\documentclass{iopart}

\usepackage{iopams} 
\usepackage{graphicx}
\usepackage{bm}
\expandafter\let\csname equation*\endcsname\relax
\expandafter\let\csname endequation*\endcsname\relax
\usepackage{amsmath}

\def\XXint#1#2#3{{\setbox0=\hbox{$#1{#2#3}{\int}$}
     \vcenter{\hbox{$#2#3$}}\kern-.5\wd0}}

\begin{document}


\title[]{Spin Excitations and Correlations in Scanning Tunneling Spectroscopy}


\author{Markus Ternes}

\address{Max-Planck Institute for Solid State Research,
Heisenbergstr. 1, 70569 Stuttgart, Germany}
\ead{m.ternes@fkf.mpg.de}

\begin{abstract}
In recent years inelastic spin-flip spectroscopy using a low-temperature 
scanning tunneling microscope has been a very successful tool for studying not 
only individual spins but also complex coupled systems. When these systems 
interact with the electrons of the supporting substrate correlated many-particle 
states can emerge, making them ideal prototypical quantum systems.
The spin systems, which can be constructed by arranging individual atoms on 
appropriate surfaces or embedded in synthesized molecular structures, can
reveal very rich spectral features. Up to now the spectral complexity has only 
been partly described. This manuscript shows that perturbation theory enables 
one to describe the tunneling transport, reproducing the differential 
conductance with surprisingly high accuracy. Well established scattering 
models, which include Kondo-like spin-spin and potential interactions, are 
expanded to enable calculation of arbitrary complex spin systems in reasonable 
time scale and the extraction of important physical properties. The emergence 
of correlations between spins and, in particular, between the localized spins 
and the supporting bath electrons are discussed and related to experimentally 
tunable parameters. These results might stimulate new experiments by providing
experimentalists with an easily applicable modeling tool.
\end{abstract}
\date{\today}


\maketitle

\section{Introduction:}

With low-temperature scanning tunneling microscopes (STM) scientist 
have developed a tool that has the ability to detect and manipulate individual 
magnetic spin systems on the atomic and molecular level by an externally 
controlled probe with sub-nm precision. These instruments have opened a new 
field of research envisioning not only a deeper understanding of the origin of 
molecular magnetism by studying the interactions between nanoscale spins but 
also of the many-particle effects between the localized spins and the itinerant 
electrons of the supporting substrate. The progress in this field is best 
demonstrated by the recently achieved ability to build stable magnetic bits by 
cleverly arranging only a handful of Fe atoms on either a thin insulating 
\cite{Loth12} or on a nonmagnetic Cu(111) surface \cite{Khajetoorians13}.
While in these experiments the transition from quantum-mechanical to classical 
magnetic behavior is explored \cite{Delgado15}, other applications might be 
envisioned ranging from spin-based logic circuits \cite{Khajetoorians11a} to 
entangled systems in which the quantum mechanical nature is crucial for 
computational purposes \cite{Leuenberger01}.

\begin{figure}[tbp]
\centering
\includegraphics[width=0.99\textwidth]{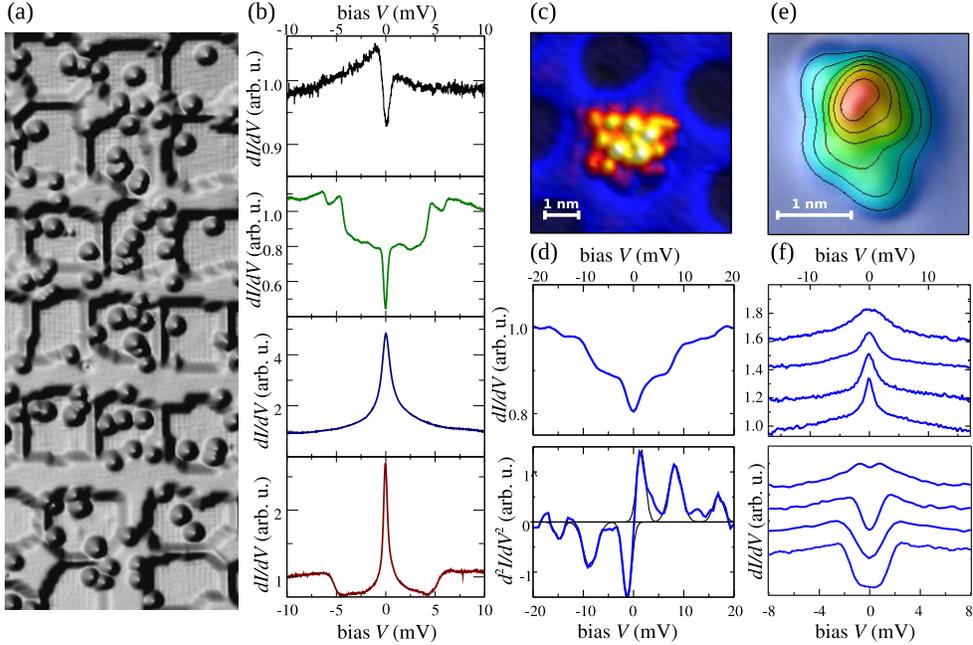}
	\caption{Some examples of inelastic spin-flip spectroscopy on single 
atoms and molecules. \textbf{(a)} Constant current topography of the Cu$_2$N 
surface on Cu(100) with co-adsorbed 3d metal atoms. (size approx.: 
$15\times40$~nm$^2$, $V=10$~mV, $I=1$~nA). \textbf{(b)} Differential 
conductance $dI/dV$ spectra measured on top of Mn, Fe, Ti, and Co adatoms (from 
top to bottom) at zero magnetic field and $T=0.6$~K. Each atom reveals a 
characteristic fingerprint in the $dI/dV$ signal. \textbf{(c)} Constant current 
topography of a Mn$_{12}$Ac$_{16}$ molecular magnet adsorbed 
on $h$-BN on Rh(111) ($V=-1$~V, $I=45$~pA). \textbf{(d)} Typical $dI/dV$ and
$d^2I/dV^2$ spectra at $T = 1.5$~K and zero magnetic field on 
Mn$_{12}$Ac$_{16}$. \textbf{(e)} 
Constant current topography of an organic radical molecule 
(C$_{28}$H$_{25}$O$_2$N$_4$) adsorbed on Au(111) ($V=100$~mV, $I=33$~pA, Contour 
line distances 50 pm).\textbf{(f)} $dI/dV$ spectra at different temperatures ($T 
= 1.5,\ 3,\ 4.4,\ 6.7$~K, bottom to top, upper panel) and magnetic fields ($B = 
14,\ 10,\ 6,\ 2$~T, bottom to top, lower panel) on the radical. Spectra are 
vertically offset for 
clarity. Data adapted from 
\cite{Otte08a,Hirjibehedin07,Kahle12, Zhang13}.}
\label{fig:Examples}
\end{figure}

At the basis of all these experiments lies the spectroscopic capability of 
the STM. About ten years ago A.~Heinrich and 
his co-workers showed in a hallmark experiment that it is possible to detect 
inelastic spin-flip excitations on Mn atoms adsorbed on patches of Al$_2$O$_3$ 
on a NiAl surface \cite{Heinrich04}. 
Since then many experiments have focused on transition metal atoms on a thin 
layer of Cu$_2$N on Cu(100) \cite{Hirjibehedin06, Hirjibehedin07, Otte08a, 
Otte09, Choi09, Ternes09, Loth10, Loth10a, Loth12, Bryant13, Oberg13, Choi14, 
Bergmann15, Spinelli14, Spinelli14z}. In these experiments, patches of Cu$_2$N 
are formed by sputtering a clean Cu(100) surface with N$^+$ ions and 
subsequent annealing. This leads to a monolayer of Cu$_2$N on the surface on 
which the desired metal atoms are deposited (Figure~\ref{fig:Examples}a). 
Experimentally, every 3d transition metal atom adsorbed on this surface reveals 
its fingerprint when a spectrum is taken by placing the tunneling tip of an STM 
over the atom (Figure~\ref{fig:Examples}b). The spectra are measured by varying 
the bias voltage applied between tip and sample and recording the differential 
conductance $dI/dV$. All spectra show characteristic features that can be 
classified into two groups: (1) Step-like increases in the differential 
conductance, positioned symmetrically around zero bias, and (2) peaks 
of the differential conductance at zero bias.  

The number of distinguishable conductance steps varies for different spin 
systems and the precise form of the steps often shows some asymmetry with 
respect to the applied bias direction and can exhibit some overshoot of 
the conductance at the step-energy. As we will see in detail, the steps are due 
to the opening of additional conductance channels precisely governed by the 
magnetic properties of the spin system. These excitations can be present even 
at zero magnetic field due to the magneto-crystalline anisotropy. The 
anisotropy is caused by the reduction of the geometric symmetry at the surface 
and by spin-orbit coupling \cite{Gambardella03, Gatteschi08, Dai08}, which have 
the effect of lifting the inherent degeneracy of the spin states. For single 
atoms the maximum anisotropy is limited to a few tens of meV \cite{Rau14}, 
whereby the magnetic excitation energies usually range from less than one meV 
up to a few meV requiring experimental setups operating at temperatures 
$T\leq4$~K. Here, it is worth to note that one has to be aware that not all 
energetically low lying step-like increases in the differential conductance 
must originate from magnetic excitations. The tunneling electrons can also 
excite low-energy mechanical vibrations which can produce similar strong 
spectroscopic features \cite{Heinrich02, Gupta05, Pivetta07, Hofer08, 
Ternes09, Natterer13a} and which can interact with the spin excitations 
\cite{Fernandez08a, May11}. Thus, to clearly distinguish magnetic excitations 
their behavior in external applied magnetic fields is often crucial. 

The peaks at zero bias are due to the Kondo effect in which 
itinerant electrons from the substrate coherently scatter with the localized 
magnetic moment of the adatom \cite{Kondo64, Kondo68, Hewson97}. This effect has 
first been detected by scanning tunneling spectroscopy on single metal atoms 
adsorbed on noble metal substrates \cite{Li98a, Madhavan98, 
Manoharan00, Madhavan01, Nagaoka02, Knorr02, Schneider02, Wahl04, Limot04, 
Wahl05}. In these early experiments the metal atoms are relatively strongly 
coupled to the substrate leading to a characteristic Kondo temperature in 
the range of $T_K\approx 30-300$~K, as determined by the full-width at 
half-maximum of the peak, and in most cases to a strongly asymmetric Fano 
lineshape due to interference effects with a potential 
scattering channel \cite{Fano61,Ujsaghy00, Kiss11}. A decoupling layer, such 
as Cu$_2$N, significantly reduces the Kondo temperature 
suggesting the possibility of influencing this many-body state with 
experimentally accessible magnetic fields and temperatures \cite{Otte08a, 
Ren14, Bergmann15, Jacobson15}. This enables the study of the 
interplay between Kondo screening, the magnetic anisotropy, and 
nearby spins \cite{Otte09, Spinelli14z}.

Apart from Cu$_2$N, other decoupling substrates have been exploited to study 
spin excitations. For example, low-lying excitations 
associated with a spin $S=1$ of the magnetic atom have been found for single 
Fe atoms embedded into the semiconducting InSb(110) surface 
\cite{Khajetoorians10, Chilian11}, which exhibits a two-dimensional electron gas 
confined to the surface. Other experiments used two-dimensional 
materials, such as graphene and hexagonal boron nitride ($h$-BN), as substrates 
for metal adatoms or magnetic molecules. For example, the Kondo state was 
observed for individual Co atoms adsorbed on a graphene sheet on top of a 
Ru(0001) surface with different Kondo temperatures and effective $g$-factors 
depending on the adsorption site with respect to the underlying metal 
\cite{Ren14}. In a different experiment Co and CoH$_x$ ($x=1-3$) complexes on 
graphene on top of Pt(111) have been investigated revealing a $S=1$ for Co and 
CoH$_3$ and a high magnetic anisotropy of $\approx8.1$~meV for the Co adsorbate 
\cite{Donati13}. On the highly corrugated and insulating $h$-BN adlayer on top 
of a Rh(111) substrate, CoH and CoH$_2$ complexes showed both, spin-flip 
excitations and the Kondo effect, pointing to two different spin states of 
$S=1$ and $S=1/2$, respectively \cite{Jacobson15}. Interestingly, the CoH 
complex revealed a dependence of the effective anisotropy on the coupling to 
the underlying substrate, similarly as observed for Co atoms on large patches 
of Cu$_2$N \cite{Oberg13}.

Spin excitations can also be observed for spin systems adsorbed on bare metal 
substrates, even though lifetime, anisotropy energies, and intensities are in 
general reduced due to the strong coupling with the substrate. For example, 
spin-excitations of Co and Fe on Pt(111) have been measured \cite{Balashov09, 
Khajetoorians13a}, where the measurements of 
reference~\cite{Balashov09} were performed at $T\approx6$~K and 
showed an approximately ten-times higher apparent magnetic anisotropy as the 
latter measurement (reference \cite{Khajetoorians13a}) that was performed at 
$T=0.3$~K. This discrepancy stems from the 
strong temperature dependent broadening of the spectrum, which leads to this 
gross overestimation and illustrates the need for low-temperature 
measurements \cite{Balashov14}. Additionally, spin excitations have been 
detected for Fe atoms adsorbed on Cu(111) \cite{Khajetoorians11} and on Ag(111) 
\cite{Chilian11a}. Apart from the 3d transition metal atoms, 4f 
lanthanide atoms also show low-energy magnetic signals. For example, the Kondo 
effect was measured on small Ce clusters on Cu$_2$N \cite{Ternes09} and 
spin-flip signals were observed for Gd adsorbed on Pt(111) and Cu(111) 
\cite{Schuh12} and Ho on Pt(111) \cite{Miyamachi13}, even though the 
intensities of the spin-flip excitations are unclear but presumably very small 
due to the strong localization of the 4f wavefunctions close to the nuclei and 
their small spacial extension \cite{Harmon74}.

In addition to these, metal-organic complexes, such as 
$M$-phthalocyanine with $M=$~Mn, 
Fe, Co, Ni, and Cu, have been studied on different surfaces 
\cite{Zhao05, Chen08, Tsukahara09, Ji10a, Franke11, Tsukahara11, Minamitani12, 
Liu13a}. Many of them showed Kondo screening with a lower Kondo temperature 
enabling the tuning by dehydrogenation \cite{Zhao05, Liu13a} or the splitting 
of the peak by an externally applied magnetic field \cite{Chen08, Tsukahara11} 
or by coupling to a ferromagnetic substrate \cite{Fu09}. Interestingly, FePc 
adsorbed on Au(111) showed a clear Kondo signature \cite{Minamitani12}, while 
the same molecule adsorbed on top of a CuO decoupling layer showed a 
double-step in the differential conductance pointing to $S=1$ 
\cite{Tsukahara09}. 

The electronic decoupling can also be created by using a substrate in the 
superconducting state in which the creation of Cooper pairs leads to a 
strongly reduced density of unpaired electrons around the 
Fermi energy. This results in a significant increase of the spin lifetime of 
the metal-organic complex \cite{Heinrich13}. Furthermore, such a 
substrate allows one to study the competition between superconducting 
phenomena and Kondo screening as observed on MnPc adsorbed on Pb(111) 
\cite{Franke11, Bauer13}.

Apart from metal atoms and metal-organic complexes spin-flip excitations 
were detected in complex molecules in which several spin centers 
are coupled forming a giant spin, as it has been shown at the example of the 
prototypical molecular magnet manganese-12-acetate-16 which has a ground state 
with a total spin of $S=10$ \cite{Kahle12} (Figure \ref{fig:Examples}c and d). 
Here, the $h$-BN decoupling layer is crucial as the magnetism is 
strongly quenched when the molecule is adsorbed upon a Au(111) surface. 
Furthermore, fully organic molecules including the charge-transfer complexes 
TTF-TCNQ (tetrathiafulvalene-tetracyanoquinodimethane) \cite{Fernandez08a} and 
an organic radical molecule \cite{Zhang13}, have shown the Kondo effect, 
whereby for the latter molecule the temperature 
and magnetic field dependence was fully understood and simulated in a 
perturbative approach (figure~\ref{fig:Examples}e and f).

To describe the inelastic spin-flip excitation spectra, scattering models have 
been developed that treat the interaction of the tunneling 
electron with the localized spin system effectively as a one-electron 
second-order perturbation \cite{Persson09, Lorente09, Gauyacq10, Hurley11}. 
Additionally, similar models allowed rationalization of the change in the 
spectra when a spin-polarized tip is employed \cite{Fransson10} and the dynamics 
at increased coupling between tip and sample \cite{Delgado10a, Delgado10, 
Hurley13}.
To address the experimentally observed bias asymmetry, co-tunneling models in 
the strong Coulomb-blockade regime have been proposed \cite{Delgado11a} and 
many-electron effects using the non-crossing approximation have been included 
\cite{Korytar12}. Furthermore, third-order scattering models similar to the 
ones used here were employed to well describe the observed bias overshooting 
at certain inelastic excitation steps \cite{Hurley11a, Hurley12}. Nevertheless, 
these models were restricted to Kondo-like interactions and did not include 
potential scattering.

In this paper, I plan to review the straightforward model of the 
exchange-interaction between an isolated spin system and the tunneling 
electrons and use a perturbative tunneling approach to simulate experimental 
data with unprecedented accuracy. The basic idea of the model goes back 50 years 
to the hallmark discovery of Juan Kondo that higher order scattering of bulk 
electrons on a magnetic impurity leads to logarithmic divergences \cite{Kondo64, 
Kondo68}. We will see that such a model enables the determination of physical 
properties like the magnetocrystalline anisotropy, the coupling strength to the 
substrate, the state lifetimes, and the magnetization directly by 
comparing the differential conductance measurements with simulations. 
Furthermore, it allows one to grasp some of the 
correlations and entanglements that are formed by the many-particle 
interactions due to the large electron bath of the substrate. The model, even 
though inherently limited due to its perturbative approach, has the advantage 
that we can develop it straightforwardly using simple matrix algebra. Due to 
this simplicity it is computationally very fast and can thus help us to 
understand future experimental observations. 
In these calculations the magnetic anisotropies, gyromagnetic factors, and 
coupling strengths to the substrate enter as experimentally determined 
parameters. Note however, that in particular for transition metal atoms 
adsorbed on metallic substrates the magnetic anisotropy have been successfully 
determined from first principle by, for example, time-dependent density 
functional theory \cite{Lounis10, Lounis11a, Khajetoorians13a, Lounis14} or 
density functional theory which includes spin-orbit coupling \cite{Shick09, 
Blonski09}.

\begin{figure}[tbp]
\centering
\includegraphics[width=0.9\textwidth]{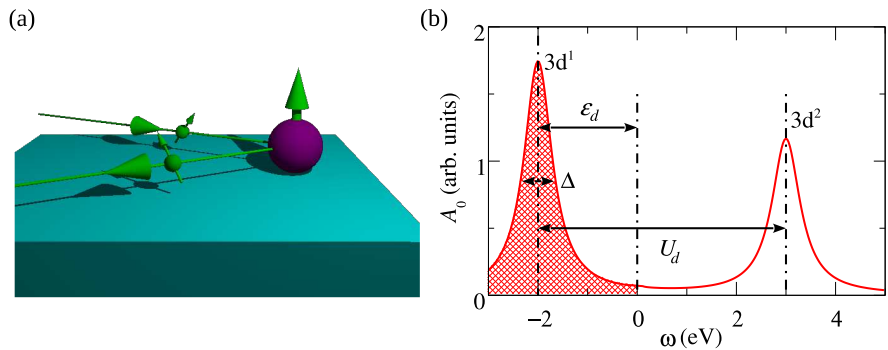}
	\caption{Comparison of the Kondo scattering and the 
single-impurity Anderson model. \textbf{(a)} 
Schematic view of the Kondo model where electrons scatter via spin-flip or the
potential interaction with a localized spin system. \textbf{(b)} 
Schematic view 
of the single orbital Anderson impurity model. An occupied 3d orbital located at 
an energy $\varepsilon_d$ below the 
Fermi energy hybridize with the substrate electrons leading to a state 
broadening $\Delta$. The occupation of this orbital with a second 
electron is prohibited by the Coulomb repulsion energy $U_d$.}
\label{fig:Schrieffer-Wolf}
\end{figure}

The basic idea of the model is to treat the tunneling as a scattering event of 
an incoming electron with the localized spin system (Figure 
\ref{fig:Schrieffer-Wolf}a). Postponing the exact meaning for later, it is 
amazing that only two parameters govern these interactions: The Kondo 
scattering parameter $J$ and the potential scattering parameter $U$. 
These two parameters are crucial for the understanding of the low-energy 
excitations. One is tempted to ask ``What is the origin of these?'' To find an 
answer it is helpful to step back from the scattering picture and look at the 
single-impurity Anderson model, which describes one half-filled atomic orbital 
(for example from a 3d transition metal atom) interacting with the continuous 
states of 
a host metal (Figure~\ref{fig:Schrieffer-Wolf}b) \cite{Anderson61}. In this 
picture all energies are in the eV range and it is not 
\emph{per se} clear how this relates to the scattering model with 
low-energy excitations in the meV range.
These relations were found by Schrieffer and Wolf, who were able to relate the 
high energies of the Anderson model with the scattering parameters 
\cite{Schrieffer66}:
\begin{equation}
J=\frac{2}{\pi}\Delta^2\left(\frac{1}{\epsilon_d}-\frac{1}{\epsilon_d+U_d}
\right) <0, \quad\quad
\mathcal{U}=\frac{2}{\pi}\Delta^2\left(\frac{1}{\epsilon_d}+\frac{1}{
\epsilon_d+U_d }
\right)
\label{equ:Schrieffer-Wolf}.
\end{equation}
Here it is crucial for the understanding of the forthcoming model that the 
Kondo spin-spin exchange scattering $J$ is for $S=1/2$ systems always 
negative, which means that the localized spin and the substrate electrons 
couple antiferromagnetically, while  the potential scattering $\mathcal{U}$ can 
have both signs, either positive or negative. However, this result can be 
different for higher spins.

The manuscript is organized as follows: In section \ref{sec:model} the 
basic model in the zero-current approximation is introduced and discussed for 
the example spin $S=1/2$. In section \ref{sec:examples} high-spin systems 
are discussed. Experimentally observed examples are 
given and compared to the model calculations. Section 
\ref{sec:limit} discusses the limitations of the model, in particular, its 
inability to cover all correlation effects that occur when a system enters the 
strong Kondo regime using experimental data obtained on the Co/Cu$_2$N system as 
an example. In section \ref{sec:Rate equations} the initial model is extended by 
including rate equations and tested against experimental data. Here, we will 
observe that  the tunneling current too can lead to the appearance of a 
non-equilibrium Kondo effect. 
Finally, section \ref{sec:summary} summarizes the manuscript and 
outlines possible routes for future extensions.

\section{The model}
\label{sec:model}
To describe the experimental observations we use a simplified model in 
which the Hamiltonian of the  total quantum mechanical system is divided into 
the ones of the subsystems of the two electron reservoirs in tip and sample, 
the one of the localized  spin system, and an interaction Hamiltonian that 
enables the exchange of charge carriers between the reservoirs:
\begin{equation}
\hat{H}=\hat{H}_t+\hat{H}_s+\hat{H}_a+\hat{H}'
\label{eq:Full_Hamiltonian}
\end{equation}
The Hamiltonian of the tip $\hat{H}_t$ and sample
$\hat{H}_s$, respectively,  can be described in the framework of second
quantization using creation and annihilation operators $\hat{a}^\dagger$ 
and $\hat{a}$:
\begin{eqnarray}
\hat{H}_t=\sum_{k,\sigma}\epsilon_{{\rm t}k\sigma}\hat{a}^\dagger_{{\rm 
t}k\sigma} \hat { a } _ {{\rm t}k\sigma } , \\
\hat{H}_s=\sum_{k,\sigma}\epsilon_{{\rm s}k\sigma}\hat{a}^\dagger_{{\rm 
s}k\sigma} \hat{a}_{{\rm s}k\sigma },
\end{eqnarray}
with $\epsilon_{{\rm t}k\sigma}$ and $\epsilon_{{\rm s}k\sigma}$ as the energy 
of the electrons with momentum $k$ and spin $\sigma$ in the tip and sample, 
respectively.

These two many-particle systems are the source and sink for the tunneling 
electrons in the STM experiment. Instead of using creation and annihilation 
operators in the momentum space $k$, we assume in the small energy range 
of interest, i.\,e. to some tens of meV around the Fermi energy, 
for tip and sample a continuous and energetically flat density of states 
$\rho(\epsilon)=\sum_{\sigma,k(\epsilon),k'(\epsilon)}\delta_{k,k'}
\left\langle \hat{a}^\dagger_{k\sigma} \hat{a}_{k'\sigma}\right\rangle 
\equiv\rho$, with $\left\langle\cdot \right\rangle$ as the time averaged 
expectation value. 

In general, the electronic states $|\varphi\rangle$ 
in these electrodes might be spin-polarized in an arbitrary direction which we 
account for by the corresponding spin density matrices 
$\varrho=|\varphi\rangle\langle\varphi|=\frac12 (\hat{I} + 
\vec{n}\cdot\hat{\boldsymbol{\sigma}})$. Here, $\vec{n}$ describes the 
direction and $-1\leq|\vec{n}|\leq 1$ the amplitude of the polarization in the 
chosen coordinate system,
$\hat{\boldsymbol{\sigma}}=(\hat{\sigma}_x,\hat{\sigma}_y,\hat{\sigma}_z)$ are
the standard Pauli matrices, and $\hat{I}$ is the $(2\times2)$ 
identity matrix. With this convention the spin polarization is 
identical to the relative imbalance between majority and minority 
spin densities
$\eta=|\vec{n}|=\left|\frac{\rho_{\uparrow}-\rho_{\downarrow}}{\rho_{\uparrow}
+\rho_ {
\downarrow}}\right|$ where $\uparrow$ and $\downarrow$ account for the two 
different spin directions along the chosen quantization axis. This description 
via density matrices allows arbitrary polarization directions in tip and sample 
which obey the quantum statistics (see \ref{ap:1}). 

The impurity spin system may either contain only a single 
spin or a finite number of coupled spins. We describe this system by a 
model Hamiltonian $\hat{H}_a$ that includes Zeeman and magnetic anisotropy 
energy and -- in the case of more than one spin -- the Heisenberg coupling 
terms $\vec{\mathcal{J}}_{ij}$\footnote{We write $\vec{\mathcal{J}}_{ij}$ as 
vector to enable also anisotropic Heisenberg or Ising-like couplings.} and the 
non-collinear Dzyaloshinskii-Moriya 
coupling $\vec{\mathcal{D}}_{ij}$ between individual spins:
\begin{eqnarray}
\fl \hat{H}_a=\sum_i g_i\mu_B\vec{B}\cdot\hat{\bf S}^i+\sum_i\left(
D_i(\hat{S}_z^i)^2+E_i\left[ (\hat{S}_x^i)^2-(\hat{S}_y^i)^2
\right] \right)\nonumber\\
+\sum_{ i ,j} \vec{\mathcal{J}}_{ij}\hat{\bf S}^i\cdot\hat{\bf S}^j+
\sum_{ i ,j} \vec{\mathcal{D}}_{ij}\hat{\bf S}^i\times\hat{\bf S}^j.
\label{eq:Atom-Hamilonian}
\end{eqnarray}
In this equation $\mu_B$ is the Bohr magneton and $g_i$, $D_i$, and 
$E_i$ are the gyromagnetic factor, the axial and the transversal 
magnetic anisotropy for the $i$-th spin in the reference coordinate frame, 
respectively.\footnote{Note, that instead of using the rather phenomenological 
$D$ and $E$ values the model can be easily adapted to more physical operators 
that connect to the spin-orbit couplings \cite{Bryant13} or to the symmetries 
of the system \cite{Miyamachi13}.} 
The externally applied magnetic field is $\vec{B}$ and the total spin 
operators $\hat{\bf S}^i=(\hat{S}_x^i, \hat{S}_y^i, \hat{S}_z^i)$ are built 
from 
operators of the form 
$\hat{S}^i_{x,y,z}=\hat{I}_1\otimes\cdots\otimes\hat{S}_{x,y,z} 
^i\otimes\cdots\otimes\hat { I } _n$, which only act on the $i$-th spin and 
where $\hat{I}$ denotes the identity matrix and $\otimes$ the Kronecker 
matrix product. The spin operators 
$\hat{S}_{x,y,z}^i$ can be easily constructed remembering that $\hat{S}_z 
|\psi_m\rangle=\hbar m_z |\psi_m\rangle$ with $m_z=S,S-1,\cdots,-S$ as the 
magnetic quantum number along our chosen $z$-axis and $\hat{S}_{\pm} 
|\psi_m\rangle=\hbar \sqrt{(S\mp m)(m\pm S+1)} |\psi_{m\pm1}\rangle$. These
then enable calculation of $\hat{S}_x=\frac{1}{2}(\hat{S}_{+}+\hat{S}_{-})$ and
$\hat{S}_y=-\frac{1}{2}i(\hat{S}_{+}-\hat{S}_{-})$, respectively. For 
simplicity we will set $\hbar=1$ from now on.

Diagonalizing the Hamiltonian of the localized spin system (equation 
\ref{eq:Atom-Hamilonian}) 
leads to discrete energy eigenvalues $\epsilon_n$ and eigenstates 
$| \psi\rangle_n$. For a single spin there are $2S+1$ eigenstates, while for 
complex spin structures that consist of several spins the number of eigenstates 
increases quickly as $\prod_i(2S_i+1)$. Because we neglect direct interactions
between the electron baths in tip, sample and the localized spin system, 
we describe the total state as a product of the continuous electron 
states $|\varphi\rangle$ and the discrete spin states $|\psi\rangle$, i.\,e.\
$|\varphi^{\rm t,s}, \psi\rangle=|\varphi^{\rm t,s}\rangle|\psi\rangle$. 
Note, however, that the coupling of the spin-system with the substrate will 
lead 
to an entanglement between sample electrons and the spin system (see section 
\ref{sec:limit}) and to a renormalization of the parameters in equation 
\ref{eq:Atom-Hamilonian} 
\cite{Wolf69, Oberg13, Delgado14, Jacobson15}. Here we 
assume that the renormalization is already included in the anisotropies and 
gyromagnetic factors and omit for the moment the entanglement.

\begin{figure}[tbp]
\centering
\includegraphics[width=0.9\textwidth]{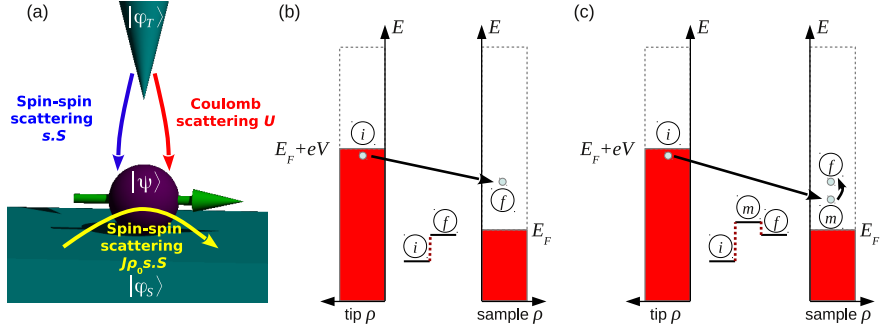}
	\caption{Model of the tunneling process between tip 
and sample. \textbf{(a)} Electrons originating from the tip interact with the 
localized spin system either via spin-spin (blue arrow) or potential scattering 
(red arrow). Additionally, we allow for spin-spin exchange scattering between 
sample electrons and the localized spin (yellow arrow). \textbf{(b)} Scheme of 
the first order Born approximation: In the tunneling process of an electron 
from the initial state $i$ to the final state $f$ it exchanges angular momentum 
and energy with the localized spin system. \textbf{(c)} In second order 
Born approximation the system additionally occupies an 
intermediate state $m$. This process produces characteristic logarithmic 
features in the tunneling spectrum. 
}
\label{fig:Model}
\end{figure}

The interaction Hamiltonian $H'$ of equation \ref{eq:Full_Hamiltonian} allows 
for tunneling of electrons from tip to 
sample or vice versa only via Kondo-like spin-flip or potential scattering 
processes with the impurity (Figure~\ref{fig:Model}a):
\begin{eqnarray}
 \hat{H}'=\hat{V}_{\rm t\rightarrow s}+\hat{V}_{\rm s\rightarrow 
t}+\hat{V}_{\rm s\rightarrow s},\nonumber\\
\hat{V}_{\rm t\rightarrow s}=
\sum_{i,\lambda',\lambda}T_{ta}^iJ_i\,\hat{a}_{{\rm 
s}\lambda'}^{\dagger}\,\hat{a}_{{\rm 
t}\lambda}\,\left(\frac12\hat{\boldsymbol{ 
\sigma}}\cdot\hat{\bf S}^i 
+U_i\hat{I}_i\right),\label{equ:Transport-Hamilonian-t-s}\\
\hat{V}_{\rm s\rightarrow t}=
\sum_{i,\lambda',\lambda}T_{ta}^iJ_i\,\hat{a}_{{\rm 
t}\lambda'}^{\dagger}\,\hat{a}_{{\rm 
s}\lambda}\,\left(\frac12\hat{\boldsymbol{ 
\sigma}}\cdot\hat{\bf S}^i
+U_i\hat{I}_i\right),\label{equ:Transport-Hamilonian-s-t}
\end{eqnarray}
with $T_{ta}^i$ as the coupling constants between tip and the 
$i$-th adsorbate spin, and $U_i=\mathcal{U}_i/J_i$ as the unitless ratio 
between Kondo and potential scattering. While we assume that the impurity is 
much more strongly coupled to the sample than to the tip, spin-spin scattering 
between the impurity and substrate electrons are additionally considered:
\begin{equation}
\hat{V}_{\rm s\rightarrow s}=\frac{1}{2}
\sum_{i,\lambda',\lambda}J_i\ \hat{a}_{{\rm 
s}\lambda'}\,\hat{a}_{{\rm 
s}\lambda}^{\dagger}\,\hat{\boldsymbol{ 
\sigma}}\cdot\hat{\bf S}^i.
\label{equ:Transport-Hamilonian-s-s}
\end{equation}

With the above assumptions the model is largely identical to the ones studied by 
Appelbaum and Anderson already in the 70's for mesoscopic tunnel junctions 
\cite{Appelbaum66, Anderson66, 
Appelbaum67}. 


\subsection{Current and conductance in second order}

We will now briefly review the calculation of the tunneling 
current using Fermi's golden rule. In STM experiments the bias $eV$ applied 
between tip and sample shifts the Fermi level $\epsilon_F$ of the tip with 
respect to the one of the sample. At positive electrons from occupied tip 
states can cross the tunnel-barrier and interact with the localized spin system 
under exchange of angular momentum and energy (Figure~\ref{fig:Model}b). 
To obey energy conservation, the energy difference 
$\epsilon_{if}=\epsilon_f-\epsilon_i$ between initial and final state of the 
spin system has to be accounted for by the energy of the tunneling electron. 
The current flowing between tip and sample is then given by:
\begin{eqnarray}
\fl
I^{\rm t\rightarrow s}(eV)= 	
\frac{2\pi e}{\hbar}T_0^2\sum_{i,f} p_i 
\int\limits_{-\infty}^{+\infty}\left|\mathcal{M}_{if}^{\rm t\rightarrow 
s}\right|^2 
f(\epsilon-eV)\left[ 
1-f(\epsilon-\epsilon_ { if } 
)\right]d\epsilon,\label{eq:current}\nonumber\\
\fl
I^{\rm s\rightarrow t}(eV)= 
\frac{2\pi e}{\hbar}T_0^2\sum_{i,f}p_i 
\int\limits_{-\infty}^{+\infty}\left|\mathcal{M}_{if}^{\rm s\rightarrow 
t}\right|^2 
f(\epsilon+eV)\left[ 
1-f(\epsilon-\epsilon_ { if } 
)\right]d\epsilon,\nonumber\\
\fl
I=I^{\rm t\rightarrow s}-I^{\rm s\rightarrow t},
\end{eqnarray}
where we have dropped the additional summation that would account for 
current through the different sites in coupled spin systems, i.\,e.\ we 
restrict our calculation to single spin systems.

The unitless tunnel barrier transmission coefficient $T_0^2=|T_{ta}J\rho_s|^2$ 
contains all 
experimental parameters that will determine the strength of the tunneling 
current. In particular, $T_0$ strongly depends on the distance between tip and 
adsorbate, and on the spin independent local densities of states in tip and 
sample. For the moment we want to assume that $T_0^2\ll1$ so that the influence 
of the tip on the spin system is negligible and the time between consecutive 
tunneling events is much longer than the relaxation time of the spin system, 
e.\,g.\ via processes such as described by 
equation~\ref{equ:Transport-Hamilonian-s-s}. Such 
conditions are usually easy to provide in STM experiments where the tip-sample 
separation can be adjusted to give tunneling currents in the pA range or below 
so that the tip can be seen as a local probe of the spin system that leaves the 
spins always in thermal equilibrium with the substrate (zero-current 
approximation).

Under this assumption the average state occupation $p_i(T)$ of the spin system 
is only governed by the effective temperature $T$ and follows the Boltzmann 
distribution,
\begin{equation}
p_i(T)=\frac{\exp[-\epsilon_i/(k_BT)]}{\sum_i 
\exp[-\epsilon_i/(k_BT)]},
\label{eq:boltzmann}
\end{equation}
with $k_B$ as the Boltzmann constant. 
Furthermore, we assume a flat density of states in tip and sample so that the 
electron occupation is given by the Fermi-Dirac distribution $f(\epsilon)=[1+ 
\exp(\epsilon/k_BT)]^{-1}$. 

While most STM experiments do not record the tunneling current directly, we are 
interested in its derivative with respect to the bias voltage. With 
the energy independent matrix elements the derivative of the current is easy 
to evaluate. Integrating the Fermi-Dirac distributions of equation 
\ref{eq:current} and calculating the derivative results in a temperature 
broadened step function \cite{Lambe68}:
\begin{equation}
\Theta(x)=\frac{1+(x-1)\exp(x)}{\exp(x)^{2}},
\label{eq:step-fkt}
\end{equation}
with $x=\epsilon/(k_BT)$, so that equation \ref{eq:current} becomes:
\begin{equation}
 \frac{{\partial} I}{\partial V} (eV) ^{\rm t\rightarrow s}= 
\frac{2\pi e^2}{\hbar}T_0^2\sum_{i,f} p_i 
\left|\mathcal{M}_{if}^{\rm t\rightarrow s}\right|^2
\Theta(eV-\epsilon_{if}). 
\label{eq:conductance}
\end{equation}

The main goal of this paper lies in deriving the transition matrix elements 
$\mathcal{M}_{if}$ to be able to solve the equation \ref{eq:conductance}. 
First, we start with second order processes, neglecting scattering that 
involves electrons which originate and end in the substrate bath, i.\,e.\ 
we concentrate on tunneling processes which consist of only one scattering event 
as depicted in figure\ \ref{fig:Model}b and described by the transport 
Hamiltonian of equations 
\ref{equ:Transport-Hamilonian-t-s} and  
\ref{equ:Transport-Hamilonian-s-t}:
\begin{equation}
\mathcal{M}^{(1)}_{if}=\left\langle\varphi_{f},\psi_{f}\right|
\frac12\hat{\boldsymbol{\sigma}}
\cdot\hat{\bf S} +U\hat{I}\left|\varphi_{i}, 
\psi_{i}\right\rangle=M_{if}+U\delta_{if}.
\label{eq:Matrix1}
\end{equation}
%
This matrix is energy independent and connects the initial states 
in the electron baths $\left|\varphi_{i}\right\rangle$ and spin system 
$\left|\psi_{i}\right\rangle$ with their final states 
$\left|\varphi_{f}\right\rangle$ and $\left|\psi_{f}\right\rangle$. It 
contains the spin-exchange scattering term $M_{if}$ and a potential scattering 
term that has non-zero matrix-elements only when initial and final spin state 
are the same ($\delta_{if}$  is the delta distribution). Computing the absolute 
square of $\mathcal{M}^{(1)}_{if}$ results in three terms that contribute to 
the tunneling conductance:
\begin{equation}
\left|\mathcal{M}^{(1)}_{if}\right|^2=\left|M_{if}
\right|^2+|U|^2\delta_{if}
+2\Re\left(U\times M_{if}\right)\delta_{if},
\label{eq:Matrix1sq}
\end{equation}
where $\Re(A)$ denotes the real part of the matrix $A$. The first term 
consist of the operators 
\begin{equation}
\hat{\boldsymbol{ 
\sigma}}\cdot\hat{\bf 
S}=\hat{\sigma}_x\hat{S}_x+ \hat{\sigma}_y\hat{S}_y 
+\hat{\sigma}_z\hat{S}_z
=
\frac{1}{2}\hat{\sigma}_+\hat{S}_-+ \frac{1}{2}\hat{\sigma}_-\hat{S}_+ 
+\hat{\sigma}_z\hat{S}_z,
\label{eq:sS}
\end{equation}
which connect the initial and final state and accounts for spin-exchange 
processes in which angular momentum between the tunneling electron and the 
localized spin system can be exchanged. The second term accounts for 
potential scattering between the tunneling electron and the localized spin 
system and 
does not change the spin state. The third term results from the interference 
between potential and spin-exchange scattering and depends, as we will 
see, strongly on the angular momentum of the  localized spin and is the origin 
of magneto-resistive tunneling \cite{Loth10, Loth10b}.

Due to the product-state of the total quantum-mechanical system, it is worth 
mentioning that the matrix elements have to be independently evaluated for the 
localized spin \textit{and} for the tunneling electron. For an electron 
tunneling between a spin-average tip and sample the non-zero matrix elements 
are:
\begin{eqnarray} 
\fl
\mbox{(a) }\langle\uparrow|\hat{s}_{+}|\downarrow\rangle=+1,\quad
\mbox{(c) }\langle\uparrow|\hat{s}_{z}|\uparrow\rangle=+1/2,
\quad
\mbox{(e) }\langle\uparrow|\hat{I}|\uparrow\rangle=+1,
\nonumber\\
\fl
\mbox{(b) }\langle\downarrow|\hat{s}_{-}|\uparrow\rangle=+1,\quad
\mbox{(d) }\langle\downarrow|\hat{s}_{z}|\downarrow\rangle=-1/2,
\quad
\mbox{(f) }\langle\downarrow|\hat{I}|\downarrow\rangle=+1,
\label{eq:arvgspinmatrix}
\end{eqnarray}
where, for completeness, we have additionally included the 
probability amplitudes for interacting with the 
unity operator $\hat{I}$. Since $\hat{\sigma}=2\hat{s}$, these 
matrix elements enable us to rewrite the spin-exchange scattering term in 
equation~\ref{eq:Matrix1sq} for a spin-average tip and sample to 
\cite{Hirjibehedin07, Loth10b}:
\begin{equation}
\fl
|M_{if}|^2= \frac{1}{2}\left|\left\langle \psi_{f}\right|
\hat{S}_-\left|\psi_{i}\right\rangle\right|^2+
\frac{1}{2}\left|\left\langle \psi_{f}\right|
\hat{S}_+\left|\psi_{i}\right\rangle\right|^2+
\left|\left\langle \psi_{f}\right|
\hat{S}_z\left|\psi_{i}\right\rangle\right|^2.
\label{eq:SA-transitionmatrix}
\end{equation}
Note, that \ref{eq:arvgspinmatrix}c and \ref{eq:arvgspinmatrix}d do not cancel
out because the two different spin directions in the initial (tip or sample) 
and final (sample or tip) bath are incoherent ensemble states and therefore 
cannot interfere with each other. 

\begin{figure}[tbp]
\centering
\includegraphics[width=\textwidth]{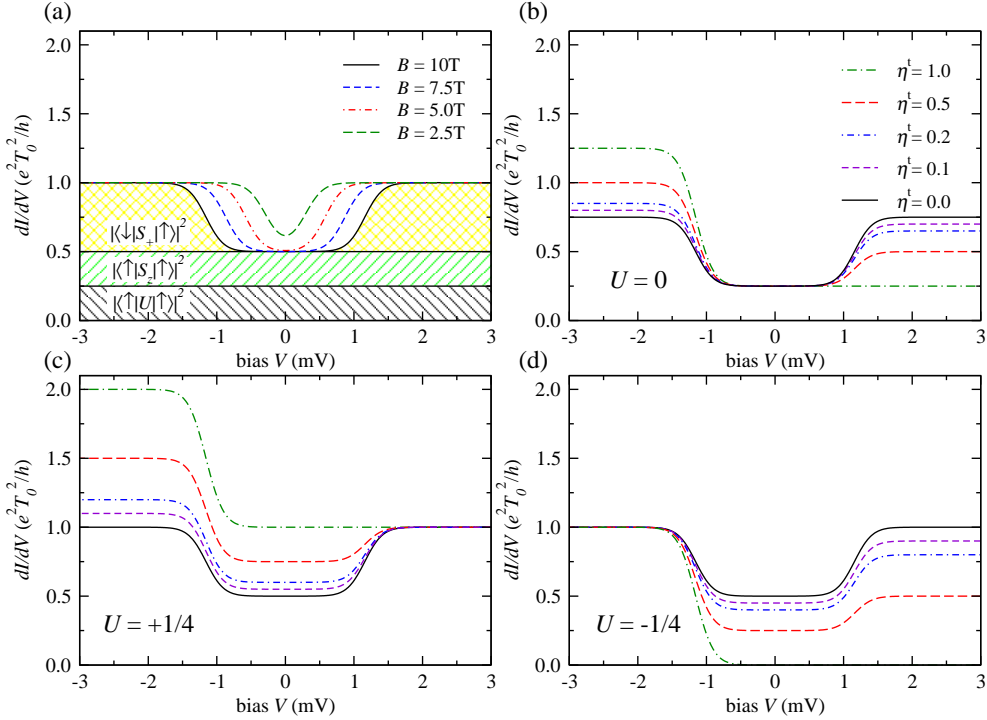}
	\caption{Tunneling spectra for a $S=1/2$ system at a temperature of 
$T=1$~K and $g=2$ evaluated in second order perturbation theory. \textbf{(a)} 
At applied magnetic field temperature broadened symmetric 
steps are visible. The different contributions to the conductance are labeled. 
Gray: conductance due to potential scattering ($|U|=0.25$), green: due 
to spin-spin scattering without changing the localized spin, yellow: 
spin-spin scattering with changing the spin. \textbf{(b)--(d)} A spin polarized 
tip  produces asymmetric steps. Here, the zero voltage conductance depends 
strongly  on the potential scattering $U$ and the tip polarization.}
\label{fig:2nd-order}
\end{figure}

For a tip with spin polarization 
$\eta=\left|\frac{\rho_{\uparrow} 
-\rho_{\downarrow}}{\rho_{\uparrow}
+\rho_{\downarrow}}\right|$ along the $z$-axis, the transition matrix 
elements are tunnel-direction dependent because either the initial or the final 
electron bath has now a spin imbalance in the density of states 
\cite{Loth10b}:
\begin{eqnarray}
\fl
|M_{if}^{\rm t\rightarrow s}|^2= \frac{1+\eta}{2}\left|\left\langle 
\psi_{f}\right|
\hat{S}_-\left|\psi_{i}\right\rangle\right|^2\!+
\frac{1-\eta}{2}\left|\left\langle \psi_{f}\right|
\hat{S}_+\left|\psi_{i}\right\rangle\right|^2\!+
\left|\left\langle \psi_{f}\right|
\hat{S}_z\left|\psi_{i}\right\rangle\right|^2\nonumber\\
\fl
|M_{if}^{\rm s\rightarrow t}|^2= \frac{1-\eta}{2}\left|\left\langle 
\psi_{f}\right|
\hat{S}_-\left|\psi_{i}\right\rangle\right|^2\!+
\frac{1+\eta}{2}\left|\left\langle \psi_{f}\right|
\hat{S}_+\left|\psi_{i}\right\rangle\right|^2\!+
\left|\left\langle \psi_{f}\right|
\hat{S}_z\left|\psi_{i}\right\rangle\right|^2.
\end{eqnarray}
As an example figure \ref{fig:2nd-order} shows the spectrum for a single 
spin with $S=1/2$. When applying the magnetic field, the conductance 
increases step-like at bias energies above the Zeeman-energy $|eV|>g\mu_B|B|$. 
These steps are symmetrical around zero voltage and thermally smeared out by 
about $5k_BT$ \cite{Lambe68}. When the electrons in the tip are spin-polarized 
the step heights at positive and negative bias are different. For a paramagnetic 
tip with a spin polarization in direction of the applied field ($\eta>0$) the 
conductance step at positive bias decreases, while it increases by the same 
amount at negative bias (Figure~\ref{fig:2nd-order}b).

With a spin-polarized tip the interference between the potential and 
Kondo-scattering shows a very particular effect \cite{Loth10, 
Loth10b}. Depending on the sign of $U$ the zero-bias conductance is either 
increased or reduced and can become even zero at $\eta=1$ and 
$U=-\frac12S$ (Figure~\ref{fig:2nd-order}c-d). This \textit{magneto-resistive 
elastic tunneling} \cite{Loth10b} arises from the third term in 
equation~\ref{eq:Matrix1sq} and scales with the expectation value $\langle 
\hat{S}_z \rangle$ of the impurity. At non-equilibrium the strength and sign of 
this term can change; this allows to read-out of the $z$-projection of the spin 
without exciting it, as has been shown in spin-pumping (see 
section~\ref{sec:Rate equations}) \cite{Loth10, 
Loth10b} and pump-probe experiments \cite{Loth10a}.

\subsection{Expansion to third order}
\label{sec:model_3}

\begin{figure}[tbp]
\centering
\includegraphics[width=0.9\textwidth]{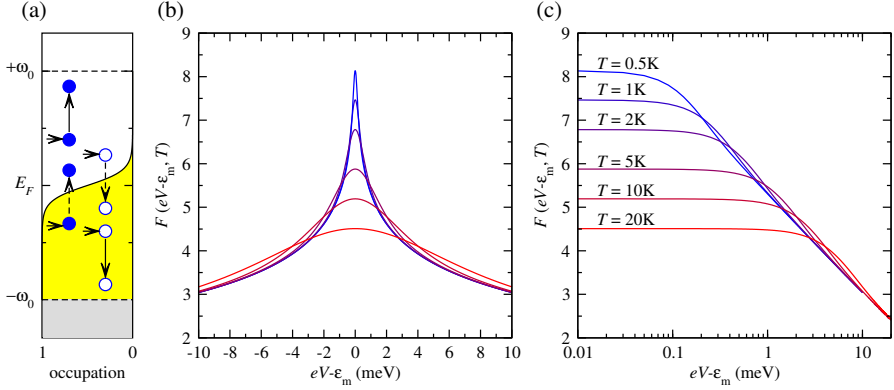}
	\caption{\textbf{(a)} Scheme of the third order processes. Full 
(dashed) vertical arrows illustrate real (virtual) transitions between 
intermediate and final state ($m\rightarrow f$) of electron-like (full circles) 
and hole-like (empty circles) carriers. \textbf{(b)} and \textbf{(c)} The 
temperature broadened logarithmic 
function  $F(eV-\epsilon_m,T)$ at different temperatures $T$ with 
$\omega_0=200$~meV and  $\Gamma_0=5$~$\mu$eV. (b) linear, (c) 
linear-log plot.}
	\label{fig:F}
\end{figure}

We now turn our attention to the next higher order of interaction. We want to 
consider processes in which, during the transport of an electron from tip to 
sample (or vice versa), the spin system additionally interacts with an electron 
from the sample bath (Figure~\ref{fig:Model}c). These processes involve an 
intermediate state ($m$) and are expressed in second order Born approximation 
as:
\begin{equation}
 \mathcal{M}^{(2)}_{if}=J\rho_s
\sum\!\!\!\!\!\!\!\!\int_m\left(\frac{\tilde{M}_{mf}M_{im}}{
\epsilon_i-\epsilon_m } +
\frac{M_{mf}\tilde{M}_{im}}{\epsilon_m-\epsilon_f}\right)\label{eq:M2}.
\end{equation} 
Here, the tilde on the matrix $\tilde{M}$ tags scattering 
processes between the localized spin and sample electrons only. The 
order of the two different electron-spin interactions can be exchanged. Thus, 
we have also to account for processes in which first an electron that 
originates and ends in the sample scatters the system into the intermediate 
state and then the tunneling electron interacts with the system scattering it 
into its final state (right fraction in equation \ref{eq:M2}). Here, it is of 
fundamental importance to have in mind that, contrary to the initial and 
final state of the total system, the intermediate state $|\psi_m\rangle$ can be 
virtual, i.\,e., must not necessarily obey angular momentum and energy  
conservation. The integro-summation symbol in equation \ref{eq:M2} indicates 
that we perform both, a summation over all possible discrete intermediate states 
in the local spin system $|\psi\rangle$ and an integration over the 
continuous states of the intermediate electron states 
$|\varphi\rangle$ in the substrate. For the integration, we consider states in 
an energy range of $\pm\omega_0$ around $E_F$ as possible scatterer leading to 
the following characteristic function for electron-like 
processes \cite{Appelbaum67, Wyatt73}:
\begin{equation}
F_e(\epsilon,T)=-\int\limits_{-\infty}^{+\infty} 
d\epsilon''\int\limits_{-\omega_0}^{+\omega_0}d\epsilon' 
\frac{1-f(\epsilon',T)}{\epsilon'-\epsilon''}f'(\epsilon''-\epsilon,T),
\label{equ:F_1}
\end{equation} 
and for hole-like processes:
\begin{equation}
F_h(\epsilon,T)=-\int\limits_{-\infty}^{+\infty} 
d\epsilon''\int\limits_{-\omega_0}^{+\omega_0}d\epsilon' 
\frac{f(\epsilon',T)}{\epsilon'-\epsilon''}f'(\epsilon''-\epsilon,T),
\label{equ:F_1h}
\end{equation} 
respectively.
Here, the Fermi-Dirac distributions in the numerator ensure an unoccupied 
(occupied) final state and the integration over the derivative $f'$ accounts for 
the temperature broadening during the total process (Figure~\ref{fig:F}a). 
The switching from virtual to real processes at $\epsilon=0$ 
leads to a logarithmic singularity that is broadened by the temperature. Thus, 
as long as the energy $\epsilon_m$ of the intermediate state and the temperature 
$k_BT$ are small compared to the integration bandwidth $\omega_0$, the 
equations \ref{equ:F_1} and \ref{equ:F_1h} (with a change of sign) can be 
rewritten as:
\begin{equation}
\fl
F(eV-\epsilon_m)\approx-\int\limits_{-\infty}^{+\infty}d\epsilon'
\ln\left(\frac{\omega_0+|eV-\epsilon_m|}{eV-\epsilon_m+i\Gamma_0}\right)
\times\Theta'(eV-\epsilon_m+\epsilon',T).
\label{equ:F_2}
\end{equation}
Here, $\Theta'(\epsilon,T)=\partial\Theta(\epsilon,T)/\partial\epsilon$ is the 
derivative of the temperature broadened step function 
(equation~\ref{eq:step-fkt}) and a small $\Gamma_0$ accounts for additional 
non-thermal (lifetime) broadening. Figure \ref{fig:F} shows the energy and 
temperature dependence of $F$ which has first been experimentally observed on a 
radical molecule adsorbed on a Au(111) surface (see figure \ref{fig:Examples}e 
and f) \cite{Zhang13}.
Compared to the Lorentzian or Frota functions \cite{Frota92} normally used 
for describing Kondo resonances in STM measurements \cite{Ternes09}, the 
behavior of this function is quite different, even though -- if analyzed only 
in a small energy range of a few $k_BT$ around zero --, it has a similar shape 
\cite{Zhang13}. The function shows a relatively flat top whose width is 
determined only by the temperature. For energies $k_BT<|\epsilon|<\omega_0$ the 
amplitude decays logarithmically as $-\ln(|\epsilon/\omega_0|)$ while the 
maximum peak height is proportional to $-\ln(k_BT/\omega_0)$. Due to the 
restriction to states in an energy interval $\pm\omega_0$ the function 
remains analytical even at $\epsilon\rightarrow\pm\infty$ where $F$ approaches 
zero. Note, that the precise value of the cut-off energy $\omega_0$ is 
not critical and mainly changes only the background offset. If not otherwise 
noted we use $\omega_0=20$~meV throughout the paper. 

When calculating the conductance we now have to consider both processes 
depicted in Figure \ref{fig:Model}b and c and thus have to replace 
$\mathcal{M}_1$ with $\mathcal{M}_1+\mathcal{M}_2$ in equation
\ref{eq:conductance} leading to:
\begin{subequations}
\label{eq:Kondo_U_M}
\begin{eqnarray} 
\fl\left|\mathcal{M}^{(1)}_{if}+\mathcal{M}^{(2)}_{if}\right|^2=&\left|\mathcal{
M } ^ { (1) } _ { if } \right|^2+\nonumber \\ 
&J\rho_s
\sum_m&\left[\left(M_{fi}\tilde{M}_{mf}M_{im}
+\overline {M_{fi}\tilde{M}_{mf}M_{im} }  
\right)F(\epsilon_{mi})+\right.\nonumber\\
&&\left.\left(M_{fi}M_{mf}\tilde{M}_{im}
+\overline {M_{fi}M_{mf}\tilde{M}_{im} }  
\right)F(\epsilon_{im})\right]\!+ 
\label{eq:Kondo-M}\\
&J\rho_sU\sum_m&\left[\left(
I_{fi}\tilde{M}_{mf}M_{im}
+\overline {I_{fi}\tilde{M}_{mf}M_{im} } \right) 
F(\epsilon_{mi})+\right.\nonumber\\
&&\left.\left(I_{fi}M_{mf}\tilde{M}_{im}
+\overline {I_{fi}M_{mf}\tilde{M}_{im} }\right) 
F(\epsilon_{im})\right]\!+\label{eq:U-M}\\
&\Or\left(J\rho_s\right)^2.\!\!\!\!\nonumber
\end{eqnarray} 
\end{subequations}
The evaluation of the matrix elements up to third order yields two 
new terms due to the interference between the processes described by 
$\mathcal{M}_1$ and $\mathcal{M}_2$ which are absent in the second order 
perturbation calculation. The term \ref{eq:Kondo-M} was first identified by 
Jun Kondo \cite{Kondo64} and can lead to temperature broadened logarithmic 
features in the conductance at the energy of intermediate states. For a 
non-vanishing potential scattering amplitude $(U\neq0)$ the term 
\ref{eq:U-M} will, in addition, produce a bias-asymmetry in the differential 
conductance even without spin-polarized electrodes. 

\begin{figure}[tbp]
\centering
\includegraphics[width=0.8\textwidth]{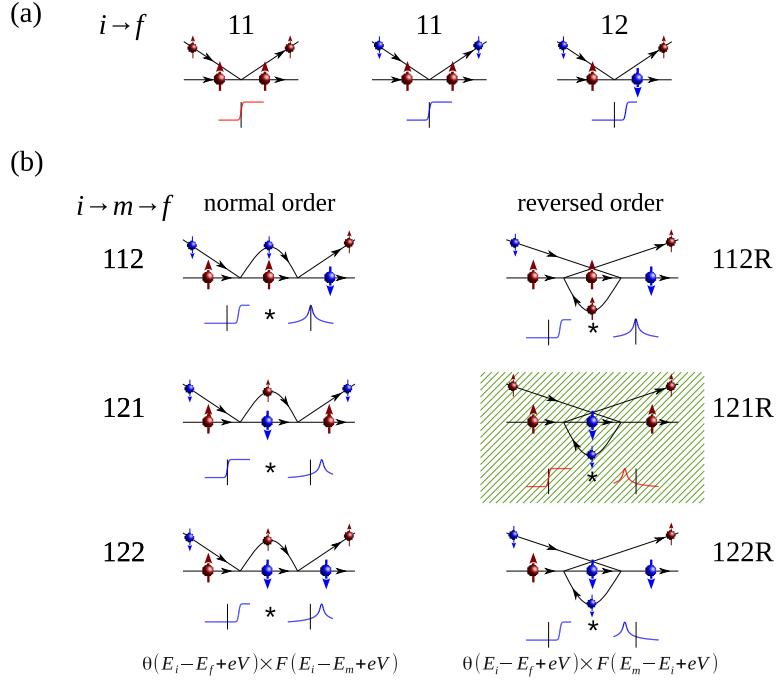}
	\caption{Interaction diagrams of order two \textbf{(a)} and three 
\textbf{(b)} for an electron tunneling from tip to 
sample into a two level $S=1/2$ spin system. The large (small) spheres 
depict the state of localized spin (interaction electron) and the color their 
spin directions. Schematic spectra show their contributions to the conductance 
at positive bias. The numbers label the processes with the state order of the 
localized spin-system. An appended 'R' label processes in which the scattering 
into the intermediate state is performed \textit{before} the tunneling electron 
interacts (exchange diagrams). Note that the time order of the processes 
influences crucially the 
conductance spectra as schematically displayed in the small graphs (vertical 
line is $E_F$; the $\ast$ means multiplication).
}
	\label{fig:Feyman}
\end{figure}

We start with the evaluation of the Kondo-like processes, that are described 
by the term \ref{eq:Kondo-M}, using a spin $S=1/2$ system with only two 
states as an example. We assume that in thermal equilibrium only the 
ground state $\left|\uparrow\right\rangle=\left|\psi_1\right\rangle$ is 
occupied, i.\,e.\ that the Zeeman-splitting induced by the external magnetic 
field is large compared to the thermal energy: $g\mu_B|B| \gg k_BT$. The 
Feynman diagrams of figure~\ref{fig:Feyman} depict all possible 
interaction processes. 
First, we review the second order processes 
(Figure~\ref{fig:Feyman}a): A spin-up electron that tunnels from tip to the 
sample cannot flip the spin ($\left|\psi_1\right\rangle\rightarrow 
\left|\psi_1\right\rangle$), while a spin-down electron can either scatter with 
exchange of angular momentum ($\left|\psi_1\right\rangle\rightarrow 
\left|\psi_2\right\rangle$) leaving the system in the state 
$\left|\downarrow\right\rangle=\left|\psi_2\right\rangle$ or without 
exchange ($\left|\psi_1\right\rangle\rightarrow \left|\psi_1\right\rangle$). 
To third order, there are a total of six diagrams to be accounted for, which 
we can label by the states occupied in the initial ($i$), intermediate 
($m$), and final ($f$) state. Here, we mark the exchange diagrams by appending a 
'R'. This results in the processes (112), (121), (122), and the 
reversed order scattering events (112R), (121R), (122R).

To evaluate the conductance due to these processes we calculate the 
spin-flip matrix elements of equation \ref{eq:sS} for the electrons and the 
localized spin-system. For electrodes without spin-polarization this results 
in conductances of:
%
\begin{eqnarray}
\fl\frac{\partial I}{\partial V}(eV)^{\rm t\rightarrow s}=\frac{4\pi 
e^2}{\hbar}T_0^2\frac{J\rho_s}{4i}\sum_{\genfrac{}{}{0pt}{}{j,k,l=}{\{x,y,z\}}} 
\varepsilon_{jkl} \Re\left[\left\langle \psi_{i}\right|
\hat{S}_l\left|\psi_{f}\right\rangle
\left\langle \psi_{f}\right|
\hat{S}_k\left|\psi_{m}\right\rangle
\left\langle \psi_{m}\right|
\hat{S}_j\left|\psi_{i}\right\rangle\right]\nonumber\\
\times F(eV-\epsilon_{im},T)
\times \Theta(eV-\epsilon_{if},T),\label{eq:3rd-normal}\\
\fl\frac{\partial I}{\partial V}(eV)^{\rm t\stackrel{R}{\rightarrow} 
s}=\frac{4\pi 
e^2}{\hbar}T_0^2\frac{J\rho_s}{4i} 
\sum_{\genfrac{}{}{0pt}{}{j,k,l=}{\{x,y,z\}}} 
\varepsilon_{jkl} \Re\left[\left\langle \psi_{i}\right|
\hat{S}_l\left|\psi_{f}\right\rangle
\left\langle \psi_{f}\right|
\hat{S}_k\left|\psi_{m}\right\rangle
\left\langle \psi_{m}\right|
\hat{S}_j\left|\psi_{i}\right\rangle\right]\nonumber\\
\times F(eV-\epsilon_{mi},T)
\times \Theta(eV-\epsilon_{if},T).\label{eq:3rd-reversed}
\end{eqnarray}
Here, equation \ref{eq:3rd-normal} accounts for the direct diagrams (normal 
order) and \ref{eq:3rd-reversed} for the exchange diagrams (reversed order), 
respectively. In these equations 
$\varepsilon_{jkl}$ is the usual Levi-Civita tensor of rank three, which is 1 
(-1) if $\{jkl\}$ is an even (odd) permutation of $\{xyz\}$, and zero otherwise 
\cite{Hurley11}. The step-function $\Theta$ ensures that energy conservation at 
the final state of the scattering process is obeyed, while the function $F$ has 
its peak at the intermediate state energy and is mirrored at $E_F$ for the 
reversed tunneling process. Any spin-polarization in tip or sample changes the 
transition matrix elements for the interacting electrons and equations 
\ref{eq:3rd-normal} and \ref{eq:3rd-reversed} would become more complex (see 
\ref{ap:1}).

\begin{figure}[tbp]
\centering
\includegraphics[width=0.8\textwidth]{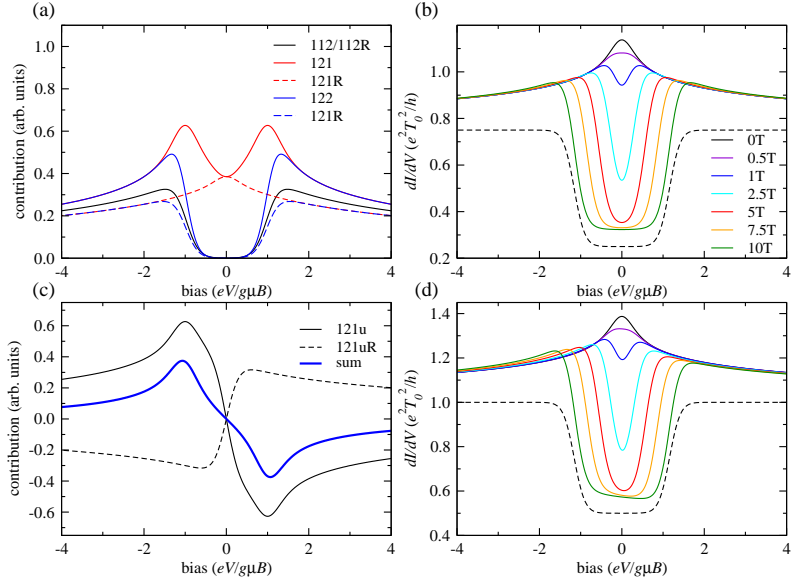}
	\caption{Tunneling spectra for a $S=1/2$ system \textbf{(a)} The third
order contributions for non spin-polarized tip and sample with $k_BT=0.1\times 
g\mu_B B$ and $\omega_0=20\times g\mu_B B$. \textbf{(b)} Total conductance for 
$J\rho_s=-0.05$, $U=0$, $T=1$~K, and $g=2$ at different magnetic fields. 
\textbf{(c)} Additional contributions to the conductance in third order due to 
a non-zero potential scattering term. \textbf{(d)} Total conductance as 
in (b) but with $U=0.25$. Dashed lines in (b) and (d) show the second order 
conductance only at $B=10$~T.}
	\label{fig:3rd-order}
\end{figure}

Comparing the different contributions to the conductance 
(figure~\ref{fig:Feyman}b and \ref{fig:3rd-order}a) it is remarkable that all 
third order scattering events start with a spin-down electron except the 
process $121R$. Here, a spin-up electron tunneling from tip to sample 
non-trivially interacts with the system because a substrate electron has 
flipped the localized spin into the intermediate spin-down state before the 
interaction takes place. 

Summing up the contributions to the conductance due to all second and 
third order scattering events results in spectra similar to those exemplary 
given in figure \ref{fig:3rd-order}b (c.\,f.\ figure 
\ref{fig:Examples}e and f \cite{Zhang13}).
At zero field the differential conductance shows a peak that splits with
increasing magnetic field. While in this calculation we assume to be in the 
weak-coupling Kondo limit and thus neglect any correlation energy due to 
the formation of a Kondo singlet, the peak splits as soon as the Zeeman energy 
overcomes the thermal energy. Note, that the resulting split-peak at small 
fields can lead to the deduction of an erroneously high $g$-factor when just 
evaluating the peak positions due to the superposition of peak and 
step-structures. 

If we consider now in addition that the scattering process between tip and 
sample or vice versa can also occur via the potential interaction (term 
\ref{eq:U-M}), we observe an asymmetric line-shape as soon as the two 
eigenstates are no longer degenerate and $\langle \hat{S_z} \rangle\neq0$
(figure~\ref{fig:3rd-order}c and d). This direction-dependent asymmetry cannot 
originate from the scattering matrix elements that involve only the 
localized spin site (the order of excitations shall be the same) but derives 
from the matrix elements involving the interacting electrons. Physically, the 
reason lies in the asymmetry of the tunnel-junction for which we assume 
that only the sample is coupled to the spin-system and thus neglect any 
intermediate scattering process that originate and end in the tip.

As an example, we examine in detail the process (121) (see 
figure~\ref{fig:Feyman}b) in both tunneling directions. For a current to flow 
from tip to sample via this process, first a $\left|\downarrow\right\rangle$ 
electron originating from the tip is scattered into a 
$\left|\uparrow\right\rangle$ state in the sample 
exciting the spin system from 
$\left|\psi_1\right\rangle\rightarrow\left|\psi_2\right\rangle$. Second, a 
$\left|\uparrow\right\rangle$ electron in the sample is scattered into 
$\left|\downarrow\right\rangle$ bringing the spin system back to 
$\left|\psi_1\right\rangle$. Third, the sample $\left|\downarrow\right\rangle$ 
is scattered back into the tip as $\left|\downarrow\right\rangle$ electron. The 
last step of this process can either take place via the $\hat{\sigma}_z$ 
or, in the case of potential scattering, via the $\hat{\sigma}_I$ term, 
respectively.  
For an electron tunneling in the reverse direction we first have a 
$\left|\downarrow\right\rangle$ from the sample being scattered into an electron 
$\left|\uparrow\right\rangle$ state of the tip simultaneously exciting the spin 
system from $\left|\psi_1\right\rangle\rightarrow\left|\psi_2\right\rangle$. 
Then, the second process flips a $\left|\uparrow\right\rangle$ sample electron 
into the $\left|\downarrow\right\rangle$ hole from the first process. Finally, 
a $\left|\uparrow\right\rangle$ tip electron traverses the junction and fills 
the $\left|\uparrow\right\rangle$ hole in the sample. 
Comparing both tunneling directions we see that it is the third process 
which differs in the initial and final state, i.\,e.\ the matrix elements are 
either $\langle\downarrow|\hat{\sigma}_{z}|\downarrow\rangle$ and 
$\langle\downarrow|\hat{\sigma}_{I}|\downarrow\rangle$ or 
$\langle\uparrow|\hat{\sigma}_{z}|\uparrow\rangle$ and 
$\langle\uparrow|\hat{\sigma}_{I}|\uparrow\rangle$. Rearranging the matrix 
elements so that all processes become electron-like, the prefactors for 
calculating the tunneling conductance due to these 
four processes are for the two without potential scattering:
\begin{eqnarray} 
(121)^{{\rm t}\rightarrow {\rm s}}:\quad
&+J\rho_s
\underbrace{\langle\downarrow^{\rm t}|\hat{\sigma}_{z}|\downarrow^{\rm 
s}\rangle}_{-}
\underbrace{\langle\downarrow^{\rm s}|\hat{\sigma}_{-}|\uparrow^{\rm 
s}\rangle}_{+}
\underbrace{\langle\uparrow^{\rm s}|\hat{\sigma}_{+}|\downarrow^{\rm 
t}\rangle}_{+}
&>0\nonumber\\
(121)^{{\rm s}\rightarrow {\rm t}}:\quad
&-J\rho_s
\underbrace{\langle\uparrow^{\rm t}|\hat{\sigma}_{+}|\downarrow^{\rm 
s}\rangle}_{+}
\underbrace{\langle\downarrow^{\rm s}|\hat{\sigma}_{-}|\uparrow^{\rm 
s}\rangle}_{+}
\underbrace{\langle\uparrow^{\rm s}|\hat{\sigma}_{z}|\uparrow^{\rm 
t}\rangle}_{+}
&>0\label{eq:sym_z},
\end{eqnarray} 
and with potential scattering:
\begin{eqnarray} 
(121u)^{{\rm t}\rightarrow {\rm s}}:\quad
&+J\rho_sU
\underbrace{\langle\downarrow^{\rm t}|\hat{\sigma}_{I}|\downarrow^{\rm 
s}\rangle}_{+}
\underbrace{\langle\downarrow^{\rm s}|\hat{\sigma}_{-}|\uparrow^{\rm 
s}\rangle}_{+}
\underbrace{\langle\uparrow^{\rm s}|\hat{\sigma}_{+}|\downarrow^{\rm 
t}\rangle}_{+}
&<0\nonumber\\
(121u)^{{\rm s}\rightarrow {\rm t}}:\quad
&-J\rho_sU
\underbrace{\langle\uparrow^{\rm t}|\hat{\sigma}_{+}|\downarrow^{\rm 
s}\rangle}_{+}
\underbrace{\langle\downarrow^{\rm s}|\hat{\sigma}_{-}|\uparrow^{\rm 
s}\rangle}_{+}
\underbrace{\langle\uparrow^{\rm s}|\hat{\sigma}_{I}|\uparrow^{\rm 
t}\rangle}_{+}&>0,
\label{eq:asym_U}
\end{eqnarray}
where we assumed $J\rho_s<0$ and $U>0$. Note, that the preceding sign change at 
the tunneling direction from sample to tip (${\rm s}\rightarrow {\rm t}$) is 
due to the rearrangement of the interaction order together with the switching 
from hole-like to electron-like scattering. The contribution to the conductance 
from the processes in which only Kondo-like spin-spin interactions take place is 
positive for both tunneling directions, while the conductance for processes that 
include potential scattering changes its sign when inverting the 
tunneling 
direction.

\section{Some single spin examples}
\label{sec:examples}

Up to here we have revised the relevant formalism to calculate the 
conductance in the third order of the matrix elements quite closely following 
the work established by Appelbaum, Anderson, and Kondo 
\cite{Kondo64, Appelbaum66, Anderson66, Appelbaum67}. 
The $S=1/2$ system we used for illustration contained only two states and was 
not influenced by any magnetic anisotropy or near-by spin systems. Recently, 
efforts have been made to expand this perturbative model to higher spin 
systems, which also include magnetic anisotropy and couplings to neighboring 
spins \cite{Hurley11a, Hurley12, Korytar12, Hurley13}, but the importance of 
the potential scattering was not taken into account up to now. 
In the following, the power of this easily accessible model will be 
used to evaluate and describe the spectral features on more complex systems. 


\subsection{The smallest ($S=$1) high-spin system}
\label{sec:S1_example}

Regarding the spectra of the $S=1/2$ system (Figure~\ref{fig:3rd-order}), 
one could get the impression that the coupling to the sample always results 
in some superimposed peak-like structures as soon as a spin flip 
transition is possible, and which scales with the substrate coupling strength 
$-J\rho_s$. In this section we will show that the situation is more 
complex and depend not only on the transition matrix elements but also on the 
state energies.

\begin{figure}[tbp]
\centering
\includegraphics[width=0.9\textwidth]{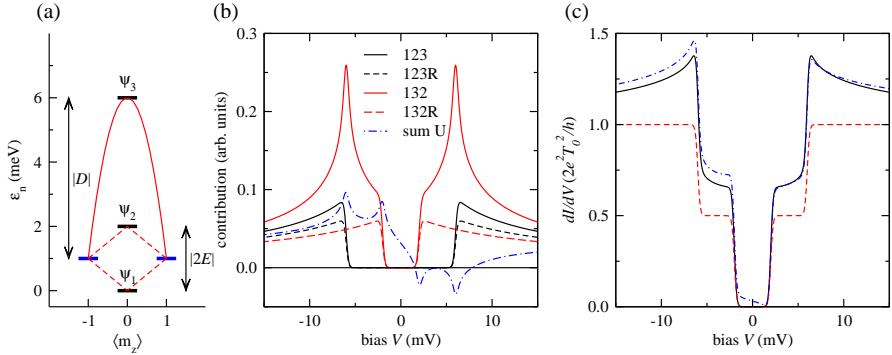}
	\caption{Tunneling spectra of a spin $S=1$ system. \textbf{(a)} The 
anisotropy lifts all degeneracies ($D=-5$~meV, $E=1$~meV) at zero field. The 
plot shows the state energy over the $m_z$ expectation value 
$\langle m_z \rangle = \tr \left( \hat{S}_z \left| \psi_i 
\right\rangle\left\langle \psi_i\right|\right)$
(blue: $E=0$, 
black: $E\neq0$) \textbf{(b)} The significant third order processes  ($T=1$~K, 
$B=0$~T). The dot 
dashed line show the additional contribution for a $U=0.125$.
\textbf{(c)} The total spectrum with $J\rho_s=0$ (dashed red line), $-0.1$ 
(full black line), and with additional $U=0.125$ (dotted dashed blue line).}
	\label{fig:Spin1}
\end{figure}

To expand the complexity, we turn now to a magnetic system with $S=1$ in which 
axial and transverse magnetic anisotropy have broken the zero-field 
degeneracy of the three eigenstates. Assuming easy-axis anisotropy ($D<0$) the 
energetically lowest eigenstates have weights only for $m_z=\pm1$. The 
additional transverse anisotropy splits the remaining two-fold degeneracy 
forming an antisymmetric ground and a symmetric first excited state 
(Figure~\ref{fig:Spin1}a). Thus, in the basis 
of the magnetic easy axis the three eigenstates can be written as: 
$\left|\psi_1\right\rangle=\frac{1}{\sqrt{2}} 
(\left|+1\right\rangle-\left|-1\right\rangle)$,
$\left|\psi_2\right\rangle=\frac{1}{\sqrt{2}}
(\left|+1\right\rangle+\left|-1\right\rangle)$, and
$\left|\psi_3\right\rangle=\left|0\right\rangle$.
Such situations have been found for example for Fe(II)-phthalocyanine molecules 
adsorbed on the ($2\times 1$) oxygen terminated Cu(110) surface 
\cite{Tsukahara09}, for Fe atoms adsorbed on InSb(110) \cite{Chilian11}, or for 
CoH complexes adsorbed on the $h$-BN/Rh(111) surface 
\cite{Jacobson15}.

By taking only second order spin-flip processes into account it is easy to 
calculate (using equation~\ref{eq:SA-transitionmatrix})
that the transition probabilities from the groundstate to the two excited 
states at $B=0$ are equal, leading to a symmetric double step structure in the 
spectrum (Figure~\ref{fig:Spin1}). Expanding the calculation to third order, 
only the processes which involve all states, i.\,e.\ (123) and (132), 
have non-zero matrix elements when the potential scattering $U=0$. 
Processes like 
(112), (121), or (131) cannot be interlinked using any combination of the 
operators $\hat{S}_+$, $\hat{S}_-$, and $\hat{S}_z$ and are thus forbidden. 
Nevertheless, additional potential scattering enables transitions involving the 
operators $\hat{S}_{+}$, $\hat{S}_{-}$, and $\hat{I}$ or $\hat{S}_{z}$, 
$\hat{S}_{z}$, and $\hat{I}$ which makes the aforementioned 
processes possible resulting in an asymmetry of the spectrum with respect 
to tunneling direction. 

It is remarkable that the remaining third order processes at $U=0$ change the 
spectrum in a quite different fashion, even though they have the same 
strength. Process (123) produces its peaks at $eV=\pm\epsilon_2$, but due 
to energy conservation it contributes to the spectrum only at 
$|eV|>\epsilon_3$, 
which efficiently cuts off the peak. In contrast, the peak at 
higher energy due to process (132) is not as strongly cut off 
(Figure~\ref{fig:Spin1}b). 
Thus, the full spectrum shows a peak-like increase of the conductance at the 
energy of the second step but not at the energetically lower first transition 
step. Furthermore, the conductance has a curved form for tunneling voltages 
between $\epsilon_2$ and $\epsilon_3$ (Figure~\ref{fig:Spin1}c). Note, the 
overall general behavior would not alter when changing from easy-axis to 
easy-plane anisotropy ($D>0$) as long as all degeneracies are lifted.

\subsection{Single Mn and Fe atoms on Cu$_2$N}
\label{sec:examplesFeMn}

\begin{figure}[tbp]
\centering
\includegraphics[width=0.8\textwidth]{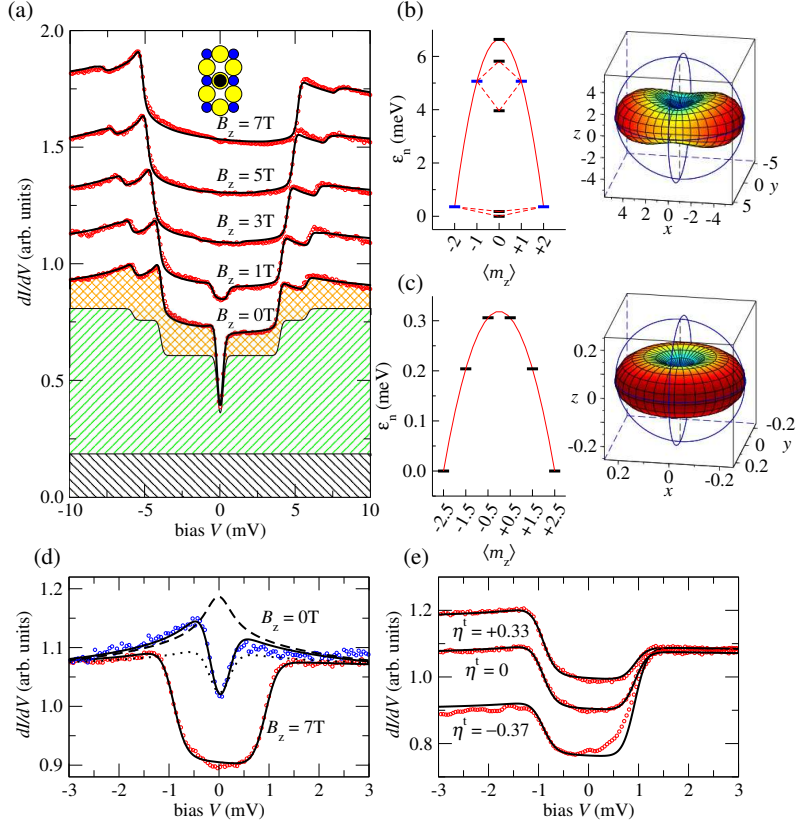}
	\caption{Comparison of experimental and calculated spectra on Fe 
and Mn atoms adsorbed on Cu$_2$N. \textbf{(a)} Experimental data from 
ref.~\cite{Hirjibehedin07} measured on a single Fe atom at increasing 
field $B_z$ along the easy axis and at a temperature $T=550$~mK (red circles). 
The simulations (black lines) for all plots are obtained with one set 
of parameters: $g=2.11$, $D=-1.57$~meV, $E=0.31$~meV, $J\rho_s=-0.087$, 
$U=0.35$, and $T_{\rm eff}=740$~mK. Additionally, a constant 
offset of $\approx20$\% of the total conductance had to been added (black 
shaded area). For $B_z=0$ the second (green 
shaded area) and third order (orange 
hatched area) contributions to the conductance are indicated. The spectra at 
field are vertically shifted for better visibility. The inset shows the 
adsorption site of the 3d atoms (black circle) on the Cu$_2$N (Cu 
yellow, N blue circle) \textbf{(b)} and \textbf{(c)} Schematic state diagrams 
and visualizations of the magnetic anisotropy (in meV) for Fe (b) and Mn (c) 
\cite{Shick09}. 
\textbf{(d)} Experimental data of two different Mn atoms at $B_z=0$ and 
$B_z=7$~T (colored circles). The fits (full lines) for the $B_z=0(7)$T data 
results in $J\rho_s=-0.029 (-0.0091)$, $U=1.35 (1.28)$, $D=-51(-39)\ \mu$eV, 
$g=1.9$, and $T_{\rm eff}=790(930)$~mK. The dashed line simulates for the zero 
field data the absence of any anisotropy. The dotted line simulates with the 7~T 
parameters the absence of a magnetic field. \textbf{(e)} The 7~T atom as in (d) 
probed with tips of different spin polarizations $\eta^{\rm t}$ (Data from 
ref.~\cite{Loth10b}).}
\label{fig:Fe}
\end{figure}

After the, with only three eigenstates, rather easy $S=1$ example, we now 
apply the model to the experimentally and theoretically intensively studied 
single 3d transition metal atoms adsorbed  on a monolayer of Cu$_2$N on 
Cu(100). When these atoms are placed on top of a Cu site they form strong 
covalent bonds with the neighboring N atoms \cite{Hirjibehedin07}. This highly 
anisotropic adsorption geometry leads to three distinct symmetry axes that are 
perpendicular to each other: The direction out-of-surface and two 
in-surface directions along the Cu-N bonds and perpendicular to it, along the 
so called vacancy rows (Figure~\ref{fig:Fe}a inset).

Single Fe atoms adsorbed on this surface have been found to be in the $S=2$ 
state with a magnetic easy-axis along the N rows ($z$-direction) and a 
magnetically hard-axis along the vacancy row ($x$-direction). In this 
coordinate system 
anisotropy values of $D=-1.55$~mV and $E=0.31$~mV, and a gyromagnetic factor of 
$g=2.11$  described the experimental data well using a spin Hamiltonian like 
equation~\ref{eq:Atom-Hamilonian} and a second order tunneling model 
\cite{Hirjibehedin07, Lorente09, Fransson09}. The Hamiltonian has as solution 
five non-degenerate eigenstates and, due to $D<0$, favors, at zero field, 
ground states with weights at high $m_z$ values. Similar to the $S=1$ system 
the transverse anisotropy breaks the degeneracies leading to a symmetric and 
antisymmetric solution with the main weights at $\left|\pm2\right\rangle$ as 
ground and first excited state and weights in $\left|\pm1\right\rangle$ for the 
second and third excited state (Figure~\ref{fig:Fe}b). 
To visualize the anisotropy we plot the total energy necessary to rotate the 
ground state into arbitrary directions showing the favored easy 
axis ($z$) and the unfavored hard axis ($x$) (Figure~\ref{fig:Fe}b) 
\cite{Shick09}. 
In second order, spin-flip scattering is allowed between the 
groundstate and the three lowest excited states but a transition to the highest 
state is forbidden because this would require an exchange of $\Delta m=\pm2$.

Experimental $dI/dV$ measurements on this system show, in addition to 
the conductance steps, peak-like structures at the second and third step but 
not at the lowest one (Figure~\ref{fig:Fe}a). Additionally, they show an 
asymmetry between positive and negative bias. To 
rationalize these observations we can follow a similar argumentation as in the 
$S=1$ case: In third order, transitions like (121) are not possible without 
additional potential interaction and processes like (123) or (124) are 
strongly cut off due to the high energy difference between $\epsilon_2$ and 
$\epsilon_{3}$ or $\epsilon_{4}$. In contrast, the processes (132) and 
(142) scale with $J\rho_s$ leading to the peak features in the differential 
conductance. 
The additional asymmetry hints at a non-negligible potential scattering. As the 
computed curves in figure \ref{fig:Fe}a reveal, our model almost perfectly fits 
the magnetic field data without any adaption of the parameters.\footnote{We 
note that all experimental data in this manuscript are unprocessed except 
for a slight slope adjustment of $<2\%$ which corresponds to the compensation 
of an unavoidable small drift in the tip sample separation of less than 
$0.5$~pm during the measurement time. The effective temperature is always larger 
than the base temperature of the experiments due to additional broadening 
effects as for example modulation voltages, electromagnetic noise, or additional 
lifetime broadening.}
We mention that similar good agreement between experimental data and 
computation can be reached in the other two magnetic field directions (not 
shown). The coupling strength in these simulations is 
$J\rho_s=-0.087$, close to the $-0.1$ found in a similar perturbative 
approach \cite{Hurley11a}.
A potential scattering term of $U=0.35$ is necessary to reproduce the 
asymmetry. This value is significantly smaller than the $U\approx 
0.75$ found in experiments where the magneto-resistive elastic 
tunneling was probed \cite{Loth10b}. Part of this discrepancy can be 
understood by an additional conductance term that does not coherently interact 
with the spin-system and which would lead to an overestimation of $U$ in 
magneto-resistive measurements. Indeed we need a constant conductance 
offset of about $20$\%, which is added to the calculated conductance to 
reproduce the spectra. 

Switching from an integer to a half-integer spin system we now discuss 
individual Mn atoms on Cu$_2$N, which have a spin of $S=5/2$ and only a small 
easy-axis anisotropy of $D\approx -40\ \mu$eV along the out-of-plane 
direction and a negligible transverse anisotropy \cite{Hirjibehedin06, 
Hirjibehedin07}. The easy-axis anisotropy prohibits the immediate formation of 
a Kondo state due to a Kramer's degenerate ground state doublet with 
$m_z=\pm5/2$ (Figure~\ref{fig:Fe}c). At zero field a typical spectrum shows 
only one step, which belongs to the transition between the $\pm5/2$ and the 
$\pm3/2$ states that have superimposed asymmetric peak structures 
(Figure~\ref{fig:Fe}d). 
The fit to the model yields $J\rho_s=-0.029$ and resembles a $S=1/2$ split-Kondo 
peak at small magnetic fields (see figure~\ref{fig:3rd-order}b). A different Mn 
atom investigated at $B_z=7$~T shows a significantly reduced $J\rho_s=-0.0091$. 
Interestingly, we find for both atoms a potential scattering value of 
$U\approx\frac{1}{2} S$, which allows one to describe the spectra without the 
need of any additional conductance offset. This high $U$ value that is the 
origin of the bias asymmetry has been independently found in spin-pumping 
experiments \cite{Loth10} 
(see section \ref{sec:Rate equations}) and 
by measuring the magneto-resistive elastic tunneling contribution 
\cite{Loth10b}. The extraordinary agreement between model and experiment can be 
seen in measurements using different spin-polarized tips on the same atom 
(Figure~\ref{fig:Fe}e). Here, the strong influence of the tip-polarization on 
the inelastic conductance  at bias voltages $|V|<1$~mV is evident while the 
differential conductance at $V>1.5$~mV stays, for all tips, constant.

\section{The limit of the perturbative approach: The Kondo system Co on 
Cu$_2$N}
\label{sec:limit}

\begin{figure}[tbp]
\centering
\includegraphics[width=1\textwidth]{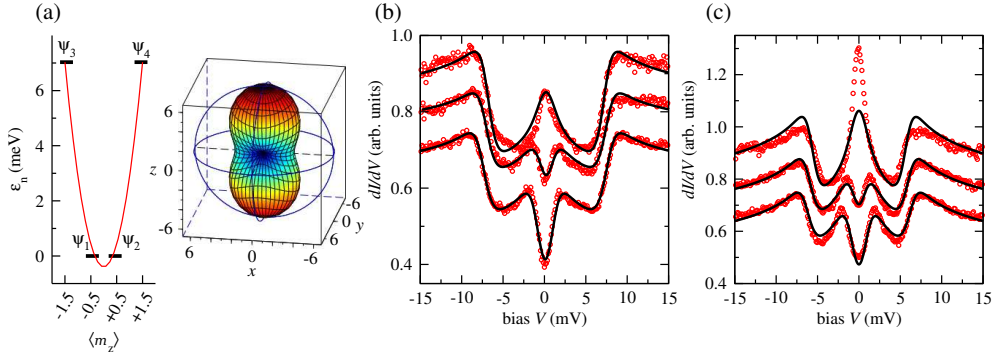}
	\caption{The tunneling spectra of Co atoms ($S=3/2$) on Cu$_2$N. 
\textbf{(a)} Schematic state diagram and visualizations of the magnetic 
anisotropy (in meV). At $B=0$ the states $|\psi\rangle_1$ and $|\psi\rangle_2$ 
are degenerated 
and differ by $\Delta m_z=\pm1$. \textbf{(b)} and \textbf{(c)} Experimental 
data from reference \cite{Oberg13} of two different Co atoms at $B_x=0,4,6$~T 
(colored circles, top to down, shifted for clarity). The best fits (full lines) 
results in $T_{\rm eff}=4$~K, $D=3.3$~meV, $E=0.7$~meV, $g=2$,
$J\rho_s=-0.11$ for (b) and $T_{\rm eff}=3.8$~K, $D=2.7$~meV, $E=0.5$~meV, 
$g=2$, $J\rho_s=-0.25$ for (c), respectively.}
\label{fig:Co-CuN}
\end{figure}

When a half-integer spin with $S>1/2$ has easy-plane anisotropy $D>0$,
its ground state at zero field is a doublet with its main weights in 
$m_z=\pm1/2$. This enables an effective scattering with the substrate electrons 
and leads at low enough temperatures to the formation of a Kondo state. 
Experimentally this has been observed for Co atoms on Cu$_2$N 
\cite{Otte08a,Choi09, Oberg13, Choi14, Bergmann15} which 
have been found to have the total spin $S=3/2$ and which enter the correlated 
Kondo state with a characteristic Kondo temperature of $T_K=2.6$~K in 
experiments performed on small patches of Cu$_2$N at temperatures down to 
$T=550$~mK \cite{Otte08a}.
Apart from $D>0$ the system also has a small in-plane anisotropy ($E\neq0$) 
which creates an easy axis ($x$) along the nitrogen row and a hard axis ($z$) 
along the vacancy rows (Figure~\ref{fig:Co-CuN}a).

Interestingly, in this system the coupling to the substrate $J\rho_s$ changes 
with the position of the Co atom on larger Cu$_2$N patches, concomitant with 
a change in the anisotropy energies which separates the 
$\left|\pm1/2\right\rangle$ states from the $\left|\pm3/2\right\rangle$ ones 
\cite{Oberg13, Delgado14}. 
For us this allows the study of the transition from the weak coupling to 
the strong coupling regime. In the case where the Co atom is relatively weakly 
coupled to the substrate ($J\rho_s\approx-0.1$), the model can be consistently 
fitted to the experimental data even at different field strengths along $B_x$ 
(Figure~\ref{fig:Co-CuN}b). We observe a zero-energy peak that splits at 
applied 
fields similar to the $S=1/2$ system of figure \ref{fig:3rd-order}d. But while 
for $S=1/2$ the field direction does not play a role, here the peak splitting 
depends strongly on the direction due to the magnetic anisotropy \cite{Otte08a}. 
For this high-spin system we furthermore observe inelastic steps due to the 
transition to energetically higher excited states which are located at 
$|eV|=2|D|$ for $B=0$ and whose additional peak structure is well described 
within the model.

For Co atoms adsorbed closer to the edges of the Cu$_2$N patches, the 
coupling to the substrate increases and the fit to the model worsens 
significantly (Figure~\ref{fig:Co-CuN}c). While the experimental data measured 
for non-zero fields are reasonable well described with $J\rho_s\approx-0.25$, 
at $B=0$ the experimentally detected peak at $E_F$ is stronger than the peak 
created by the model. 
Additionally, the experimental peak-width appears to be smaller than the 
temperature broadened logarithmic function, which was similarly observed for a 
radical molecule with $S=1/2$ for temperatures presumably close to $T_K$ 
\cite{Zhang13}. Furthermore, we observe that the calculated spectrum no 
longer well describes the steps at $2|D|$. At this energy the third 
order contributions are less pronounced in the experimental data, indicating 
that we reach the limit of the perturbative approach.

\begin{figure}[tbp]
\centering
\includegraphics[width=0.95\textwidth]{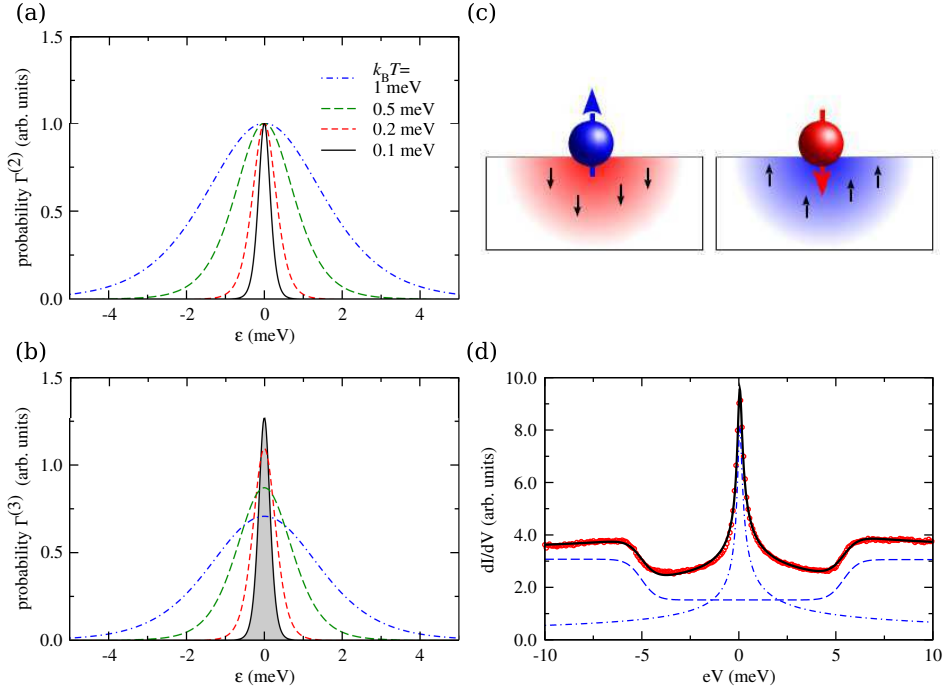}
	\caption{Correlations induced by substrate electrons.
\textbf{(a)} In second order the scattering probability of a substrate 
electron with energy $\epsilon$ is given by the overlap of the electron 
and hole-like Fermi-Dirac distributions (area underneath the curve) and is for 
degenerated ground states directly proportional to the temperature. Third order 
scattering \textbf{(b)} gets logarithmically stronger than 
the second order processes with decreasing 
temperature. \textbf{(c)} Schematics of the entangled state at 
temperatures below the characteristic Kondo 
temperature. The scattering leads to a correlation between the spin state of 
the 
localized system and the sample electrons in the vicinity. \textbf{(d)} 
Spectrum of a Co adatom on Cu$_2$N in the Kondo regime ($B=0$~T, 
$T=0.55$~K, data from ref.~\cite{Otte08a}). The experimental $dI/dV$ curve (red 
dots) is well described by the sum (full line) of a Frota peak centered at 
$E_F$ (dotted dashed line) and a broadened step-function at higher energy 
(dashed line).}
\label{fig:Sample_entanglement}
\end{figure}

The full description of a spin system in the strong coupling regime 
requires complex theoretical methods like numerical renormalization group theory 
\cite{Wilson75, Hewson97, Bulla08} which are beyond the scope of this paper. 
Nevertheless, we can discuss some of the physical consequences within the 
outlined perturbative model. In contrast to the examples discussed in section 
\ref{sec:examples}, the two lowest degenerate ground states of the Co/Cu$_2$N 
system have weights in states that are separated by $\Delta m=\pm1$. Thus, at 
zero field, electrons from the substrate can efficiently flip between these two 
states. The computation of the transition rate between the two states 
$|\psi_i\rangle$ and $|\psi_f\rangle$ of the spin system that have the energies 
$\epsilon_i$ and $\epsilon_f$ is $\Gamma_{if}:=I_{if}/e$, i.\,e.\ the transition 
probability per time, is analogous to the calculation of the current (equation 
\ref{eq:current}) given by:
\begin{equation}
\Gamma_{if}^{\rm s\rightarrow s}=\frac{I_{if}^{\rm s\rightarrow s} }{e}
=\frac{2\pi}{\hbar}(J\rho_s)^2\int_{-\infty }^{ \infty}d\epsilon\, 
|\mathcal{M}_{if}^{\rm s\rightarrow 
s}|^2 f(\epsilon,T)[1-f(\epsilon-\epsilon_{if},T)].
\label{equ:Gamma_SS}
\end{equation}
Here we have set the bias to $eV=0$ and replaced the coupling constant 
$T_0$ between adsorbate and tip with $J\rho_s$, the coupling between 
adsorbate and substrate. When we now consider only processes up to second 
order, the matrix $\mathcal{M}_{if}$ is independent of the energy and the 
integral can be evaluated to
\begin{equation}
\int_{-\infty }^{ \infty}d\epsilon\, 
f(\epsilon,T)[1-f(\epsilon-\epsilon_{if},T)]=\frac{\epsilon_ { if}}{
\exp\left(\frac{\epsilon_ { if}}{k_BT}\right)-1}.
\label{equ:box}
\end{equation}
Thus, the scattering rate between the two degenerate states 
(i.\,e.~$\epsilon_{12}=0$) is directly proportional to the temperature 
(Figure~\ref{fig:Sample_entanglement}a):
\begin{equation}
\Gamma_{12}^{(2)\rm s\rightarrow 
s}=\frac{2\pi}{\hbar}(J\rho_s)^2|\tilde{M}_{12}|^2k_BT.
\label{equ:Gamma_SS(2)}
\end{equation}

These scattering processes, which we have discussed in section~\ref{sec:model} 
and displayed in figure~\ref{fig:Feyman}a, will tend to change the spin 
polarization of the electronic states in the sample \emph{near the adsorbate} to 
be correlated with the localized spin. Nevertheless, this local correlation will 
be quickly destroyed by decoherent scattering processes with the remaining 
electron bath, which we can assume to be large and dissipative. This 
decoherence rate is also proportional to the temperature, 
$\Gamma_{decoh}\propto 
k_BT$ \cite{Delgado12}, but usually stronger so that no highly correlated state 
can form. This picture changes when we additionally
consider third order scattering processes which yield, using 
equation \ref{eq:Kondo_U_M}, the probability:
%
\begin{eqnarray}
\fl\Gamma_{if}^{(3)\rm s\rightarrow 
s}=\frac{4\pi}{\hbar}(J\rho_s)^3\int_{-\infty }^{ \infty}d\epsilon\,\sum_m 
\Re(\tilde{M}_{fi}\tilde{M}_{mf}\tilde{M}_{im})\nonumber\\
\times\left[F(\epsilon_{mi}-\epsilon,T)+F(\epsilon_{im}-\epsilon,T)\right]
f(\epsilon, T) [
1-f(\epsilon-\epsilon_ { if }, T)].
\label{equ:Gamma_SS(3)}
\end{eqnarray}

Due to the growing intensity of $F(\epsilon= 0)$ at reduced temperatures 
(Figure~\ref{fig:F}), for temperatures $T\rightarrow 0$ the scattering 
$\Gamma^{(3)}$ decreases significantly more slowly than $\Gamma^{(2)}$ (see 
figure~\ref{fig:Sample_entanglement}b) so that their ratio steadily increases:
\begin{equation}
\frac{\Gamma^{(3)}}{\Gamma^{(2)}}\approx 
J\rho_s\ln\left(\frac{k_BT}{\omega_0}\right).
\label{equ:Gamma3_SS}
\end{equation}
In contrast to that, the decoherent scattering rate  
with the bath $\Gamma_{decoh}$ lacks localized scattering centers and 
therefore has no significant third order 
contributions. Equation \ref{equ:Gamma3_SS} leads to a characteristic 
temperature, the Kondo temperature $T_K$, where $\Gamma^{(3)}$ and 
higher order scattering terms become the dominant processes and 
perturbation theory breaks down \cite{Kondo64, Suhl65, Nagaoka65, Hewson97}:
\begin{equation}
T_K\approx \frac{\omega_0}{k_B}\exp\left(\frac{1}{J\rho_s}\right).
\label{equ:TK}
\end{equation}
Below this temperature the assumption used up to now, i.\,e.\ that the 
electronic states in the sample are not influenced by the presence of the spin 
system, no longer applies. Using exact methods like the modified Bethe ansatz 
\cite{Andrei80, Wiegmann80} or numerical renormalization group theory 
\cite{Bulla08} reveals that the sample electrons in the vicinity rather form an 
entangled state with the impurity, i.\,e.\ 
\begin{equation}
\Psi^{total}=\frac{1}{\sqrt{2}} 
\left(\left|\downarrow^{\rm s}\right\rangle\left|+\frac12\right\rangle-
\left|\uparrow^{\rm s}\right\rangle\left|-\frac12
\right\rangle\right),
\label{equ:Kondo-state}
\end{equation}
as illustrated in 
figure~\ref{fig:Sample_entanglement}c. This combined state is quite 
complicated because the electronic states in the sample are continuous in energy 
and extend spatially, and therefore strongly alter the excitation spectrum of 
the adsorbate \cite{Costi00, Zitko09}. 

Figure \ref{fig:Sample_entanglement}d shows the impact of the formation of the
correlated state on the experimentally detected spectrum of the 
Co/Cu$_2$N system. At temperatures $T\ll T_K$, the 
peak at the Fermi energy can no longer be well reproduced by a temperature 
broadened logarithmic function (which in any case must diverge for 
$T\rightarrow 0$). The peak is much better described by an asymmetric 
Lorentzian, i.\,e.\ a Fano line-shape \cite{Fano61}, or the Frota function 
\cite{Frota92, Prueser11, Zitko11a} which has a finite amplitude and a 
half-width of $\Delta=k_BT_K$, corresponding to the correlation energy of the 
Kondo state. Superimposed on this zero-energy peak we detect conductance steps 
at voltages that enable scattering of the spin system into the $m_z=\pm3/2$ 
state. These steps are well described using only a second order perturbation 
spin-flip model \cite{Otte08a} and do not show any third order logarithmic 
contributions. This means that the probability of the third order scattering 
channels must be closed due to the ground-state correlation between the 
localized spin and the substrate electrons. Here we see that the simple 
perturbative approach of the model is no longer valid and fails to capture the 
full physics in this strongly coupled Kondo regime. Nevertheless, the model 
still gives valuable information because it enables us to identify the 
conditions under which correlations can evolve and is due to its computational 
simplicity also applicable for larger and more complex spin systems where the 
calculation of the exact solution might not be feasible or very time consuming.


\section{Spin dynamics and rate equations}
\label{sec:Rate equations}

In the preceding section we discussed the appearance of correlations due to 
higher order scattering between the substrate electrons and the localized spin 
system. Now we want to evaluate how the tunneling of electrons 
between the biased tip and the sample influences the state populations and the 
observable spectroscopic features in local conductance measurements. This is 
equivalent to dropping the zero-current approximation we have assumed until now. 
We will see that the change of time-averaged system properties results in 
characteristic fingerprints in the $dI/dV$ signal which can become crucial for a 
deeper understanding of the system's response and dynamics.

To describe the transition rate between the localized spin states 
$\left|\psi_i\right\rangle$ and $\left|\psi_f\right\rangle$ due to the 
tunneling current flowing between tip and sample we 
can again use equation \ref{eq:current} and the relation $\Gamma_{if}^{\rm 
t\rightarrow s}(eV):=I_{if}^{{\rm t}\rightarrow{\rm s}}(eV)/e$, which leads, in 
first order Born approximation, to:
\begin{equation}
\Gamma_{if}^{(2)\rm 
t\rightarrow s}(eV)=\frac{2\pi}{\hbar}T_0^2\int_{-\infty }^{ 
\infty}d\epsilon\, 
|\mathcal{M}_{if}^{\rm t\rightarrow 
s}|^2 f(\epsilon-eV,T)[1-f(\epsilon-\epsilon_{if},T)].
\label{equ:Gamma_TS1}
\end{equation}
The integral can be solved using equation~\ref{equ:box} resulting in:
\begin{equation}
\Gamma_{if}^{(2)\rm 
t\rightarrow s}(eV)=\frac{2\pi}{\hbar}
T_0^2 |\mathcal{M}_{if}^{\rm t\rightarrow s}|^2 
\frac{\epsilon_ { 
if}-eV}{\exp\left(\frac{\epsilon_ { 
if}-eV}{k_BT}\right)-1}.
\label{equ:Gamma_TS2}
\end{equation}
As long as $eV-\epsilon_{if}\gg k_BT$, equation~\ref{equ:Gamma_TS2} can be 
further 
simplified to an equation that is linear in the energy difference:
\begin{equation}
\Gamma_{if}^{(2)\rm 
t\rightarrow s}(eV)=(2\pi/\hbar)
T_0^2 |\mathcal{M}_{if}^{\rm t\rightarrow s}|^2 (eV-\epsilon_{if}).
\end{equation}
This linear dependence of the scattering rate on the energy difference is 
equivalent to the assumptions that in second order perturbation the matrix 
elements and the coupling constant $T_0$ between tip and sample are energy 
independent. 
Under these assumptions the rate is given by the energy window between the 
available energetically hot electrons and the energy needed to change the 
localized state \cite{Delgado10a}. However, one should keep in mind that 
at large energy differences the intrinsic variation of the local density of 
electronic states in tip and sample $\rho(\epsilon)^{\rm t, s}$ might alter the 
results. 

Furthermore, we additionally include third order contributions to the tunneling 
transition rates. This can be archived by integrating the interference 
terms between $\mathcal{M}_1$ and $\mathcal{M}_2$ of equation~\ref{eq:Kondo_U_M} 
and leads to an additional contribution to the scattering rate of:
\begin{eqnarray}
\fl\Gamma_{if}^{(3)\rm t\rightarrow s}(eV) 
=\frac{4\pi}{\hbar}T_0^2J\rho_s\int_{-\infty }^{ \infty}d\epsilon\,\sum_m
f(\epsilon-eV, T) [
1-f(\epsilon-\epsilon_ { if }, T)]\times\nonumber\\
\left[\Re(M_{fi}\tilde{M}_{mf}M_{im})F(\epsilon_{mi}
-\epsilon,T)\right.+\nonumber\\
\left.\Re(M_{fi}M_{mf}\tilde{M}_{im})F(\epsilon_{im}
-\epsilon,T)\right.+\nonumber\\
\left.U\Re(I_{fi}\tilde{M}_{mf}M_{im})F(\epsilon_{mi}
-\epsilon,T)\right.+\nonumber\\
\left.U\Re(I_{fi}M_{mf}\tilde{M}_{im})F(\epsilon_{im}
-\epsilon,T)\right].
\label{equ:Gamma_TS3}
\end{eqnarray}
For tunneling currents flowing in the opposite direction, i.\,e.\ for 
electrons which are scattered at the spin system when tunneling from sample to 
tip, equations \ref{equ:Gamma_TS2} and \ref{equ:Gamma_TS3} can be 
adapted straightforwardly. 

The total transition rates $\Gamma_{ij}^{\rm t\rightarrow 
s}=\Gamma_{ij}^{(2)\rm t\rightarrow s}+\Gamma_{ij}^{(3)\rm t\rightarrow s}$ 
and $\Gamma_{ij}^{\rm s\rightarrow 
t}=\Gamma_{ij}^{(2)\rm s\rightarrow t}+\Gamma_{ij}^{(3)\rm s\rightarrow t}$ 
have to be evaluated for all possible initial and final localized spin states 
and, together with the dissipative substrate to substrate scattering rates 
discussed  in section \ref{sec:limit} (equations \ref{equ:Gamma_SS} and 
\ref{equ:Gamma_SS(3)}), can be brought together to the characteristic master 
equation for the state populations:
\begin{eqnarray}
\fl\frac{d}{dt}p_i(t)=\sum_{j\neq i}p_j(t)(\Gamma_{ji}^{\rm t\rightarrow 
s}+\Gamma_{ji}^{\rm s\rightarrow t}+\Gamma_{ji}^{\rm s\rightarrow s})-
p_i(t)\sum_{j\neq i}(\Gamma_{ij}^{\rm t\rightarrow 
s}+\Gamma_{ij}^{\rm s\rightarrow t}+\Gamma_{ij}^{\rm s\rightarrow 
s})\nonumber\\
{}=\Upsilon_{ij}(t)p_j(t).
\label{equ:Master}
\end{eqnarray}
This set of first order differential equations describes the time-development 
of an initial state of the spin system, which might be a superposition state 
$\left|\psi(t)\right\rangle=\sum_ip_i(t)\left|\psi_i\right\rangle$, at the 
time $t$ under the influence of all possible, bias dependent, scattering 
events. The first summation in equation~\ref{equ:Master} accounts for all the 
probabilities to scatter into the $i$-th eigenstate from any other eigenstate. 
These probabilities have to be weighted with the time-dependent state 
probability $p_j$ of the originating state. The second summation accounts for 
all scattering events 
which reduces the $i$-th state probability by scattering it into 
other eigenstates. It is important to note, that we use three crucial premisses 
in this approach: 
\begin{enumerate}
 \item We assume that the future of the localized quantum 
state depends only on the actual state and not on any interactions which 
happened in the past. This Markovian limit means that we assume that the 
timescales at which this approximation breaks down, in particular where 
higher than two-time correlations are dominant, can be neglected 
\cite{Cohen89}. 


\item We do not account for any phase coherences between the eigenstates 
of the spin system. This means that the phase coherence time is short 
compared to the state's lifetime $\tau$ and the time between 
successive tunneling events. Under these assumptions all off-diagonal elements 
of the density matrix in the rotating frame with the eigenstates $\psi'$ are 
zero and thus the density matrix can be described as a state 
mixture $\chi=\sum_i p_i \left|\psi'_i\right\rangle\left\langle\psi'_i\right|$.

\item Because we treat the total system as the product state between the 
continuous electronic states in tip and sample and the discrete spin states, a 
correlated state between substrate electrons and the localized spin as 
discussed in section~\ref{sec:limit} (equation~\ref{equ:Kondo-state}) can not 
be developed directly within the limitations of our model.
\end{enumerate}

In the following, we further want to limit our evaluation to steady-state 
conditions, which means that the change of any external parameter, like the 
tunneling bias voltage, occurs adiabatically slowly, much more slowly than any 
relaxation times in the system.
The steady-state condition is reached, when all $p_i$ are static, i.\,e.\ 
$\frac{d}{dt}p_i(t\rightarrow\infty)=0$. This is equivalent to finding of
the algebraic kernel of the rate matrix $\Upsilon$, which additionally has to 
be normalized to account for the conservation of probabilities 
$\sum_ip_i=1$:
\begin{equation}
p^{stat}(eV)=\frac{\ker \Upsilon(eV)}{\|\ker \Upsilon(eV)\|}.
\label{eq:p_stat}
\end{equation}
From these $p^{stat}(eV)$ the current $I(eV)$ and the differential conductance 
$dI(eV)/d(eV)$ can be calculated:
\begin{equation}
I(eV)=e\sum_{i,j} p_i^{stat}(eV) \left(\Gamma_{ij}^{\rm t\rightarrow s}(eV)
-\Gamma_{ij}^{\rm s\rightarrow t}(eV)\right).
\end{equation}

\begin{figure}[tbp]
\centering
\includegraphics[width=\textwidth]{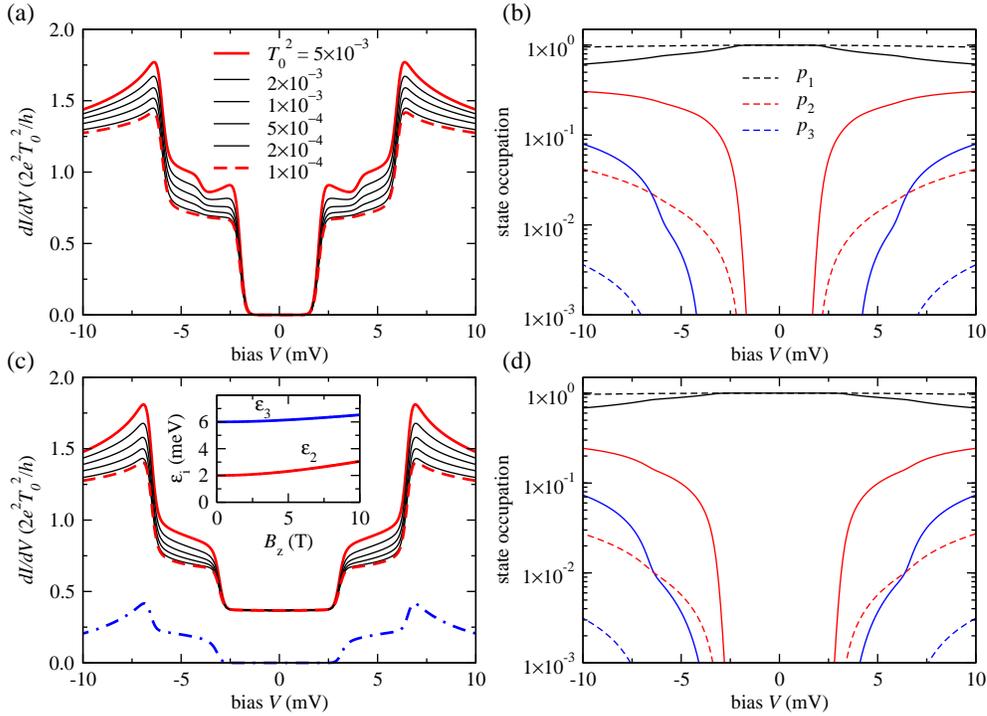}
	\caption{Tunneling spectra of a spin $S=1$ system at 
different coupling strengths $T_0^2$ between tip and sample. The simulation 
parameters are identical to the ones used in figure~\ref{fig:Spin1}: $g=2$, 
$D=-5$~meV, $E=1$~meV, $U=0$, $J\rho_s=-0.1$, $T=1$~K.  \textbf{(a)} Normalized 
zero field spectra for $T_0^2$ ranging from $1\times 10^{-4}$ (dashed red line) 
to $5\times 10^{-3}$ (full red line). At increased $T_0$ a feature at 
$eV=\pm|\epsilon_3-\epsilon_2|\approx \pm 4$~meV appears and the apparent 
strength of the peaks at $eV=\pm\epsilon_3$ increase. \textbf{(b)} 
State occupancy in steady-state for the smallest (dashed lines) and highest 
(full lines) coupling strengths. \textbf{(c)} and \textbf{(d)}  Same as (a) and 
(b) but at an applied field of $B_z=10$~T. The dashed-dotted line in (c) is the 
relative difference between the spectra at highest and lowest coupling strength. 
The inset in (c) shows the evaluation of the state energy 
$\epsilon_i$ with field. 
}
\label{fig:Pumping_S1}
\end{figure}

To show the influence of a non-zero tip-sample coupling strength $T_0$ on the 
tunneling spectra we are returning to the spin $S=1$ example from 
section~\ref{sec:S1_example}. In figure~\ref{fig:Pumping_S1}a and c we show 
the development of the $dI/dV$ spectra at $B=0$ and $B_z=10$~T for different 
coupling strengths between sample and tip, respectively. 
The coupling constants correspond to a tunneling resistance between 
$R_T\approx130$~M$\Omega$ and $\approx2.6$~M$\Omega$, which is equivalent to a 
tunneling current of $I\approx 150$~pA -- $8$~nA at a bias of $V=20$~mV. Note, 
that these are quite typical parameters for STM experiments. 

Without magnetic field and at small couplings 
we observe a spectrum which does not differ significantly from the one 
calculated in the zero-current approximation. The average occupation of the 
two excited states increases only slowly for voltages above the 
threshold for these transitions and reaches at $V=\pm10$~mV only 
about $4\%$ 
for $p_2$ and less than $0.4\%$ for $p_3$ (dashed lines in 
figure~\ref{fig:Pumping_S1}b). This situation changes when the tip-sample 
coupling is increased. At larger $T_0$ an additional feature appears in the 
spectrum at about $\pm4$~mV, which is due to transitions from the state 
$\left|\psi_2\right\rangle$ to $\left|\psi_3\right\rangle$. These transitions 
are only possible because the probabilities $p_i$ are driven far from 
thermal equilibrium and $p_2$ has already a significant weight at 
$|V|=4$~mV. Similar current induced pumping to higher excitation states 
has been observed for example for Mn-dimers adsorbed on Cu$_2$N 
\cite{Loth10}, for small Fe clusters containing only a few atoms on Cu(111) 
\cite{Khajetoorians13}, or for Fe-OEP-Cl (Fe-octaethylporphyrin-chloride) 
molecules adsorbed on Pb(111) \cite{Heinrich13a}. Interestingly, the latter 
experiment was performed on a superconducting surface and with a superconducting 
Pb-tip which made it necessary to account for the gap around the Fermi energy in 
the quasi-particle density of states of tip and sample.

If we now add a magnetic field of $B_z=10$~T to our simulation , the 
situation changes: The energy differences $\epsilon_{12}$ and $\epsilon_{23}$ 
between the first and second, and second and third eigenstate, respectively, 
are almost identical (see figure~\ref{fig:Pumping_S1}c inset) masking the 
additional step-like feature when pumping into energetically higher states. 
Nevertheless, also here a clear change of the spectrum occurs at high tip-sample 
couplings. We observe an increased relative conductance between $\epsilon_2$ and 
$\epsilon_3$ and a stronger third order peak at $eV=\pm\epsilon_3$. This feature 
is mainly due to the third order scattering process $(232)$ now possible, as 
apparently visible when taking the relative difference between the spectra at 
strongest and weakest coupling strength (dashed dotted line in 
figure~\ref{fig:Pumping_S1}c).

As long as the tip is not spin-polarized, the current induced pumping into 
higher states cannot lead to an inversion of the state occupancy. The 
state's lifetime $\tau_i$ for the eigenstate $\left|\psi_i\right\rangle$ is 
inversely proportional to the sum of all scattering processes which leaves the 
state:
\begin{equation}
\tau_i^{-1}=\sum_{j\neq i}\left(\Gamma_{ij}^{\rm 
t\rightarrow 
s}(eV)+\Gamma_{ij}^{\rm s\rightarrow t}(eV)+\Gamma_{ij}^{\rm s\rightarrow,
s}(eV)\right).
\label{eq:T1}
\end{equation}
While for spin-averaged electrodes the scattering matrix elements 
do not change when inverting the initial and final state, the energetically 
higher states must have always a shorter lifetime.

\begin{figure}[tbp]
\centering
\includegraphics[width=\textwidth]{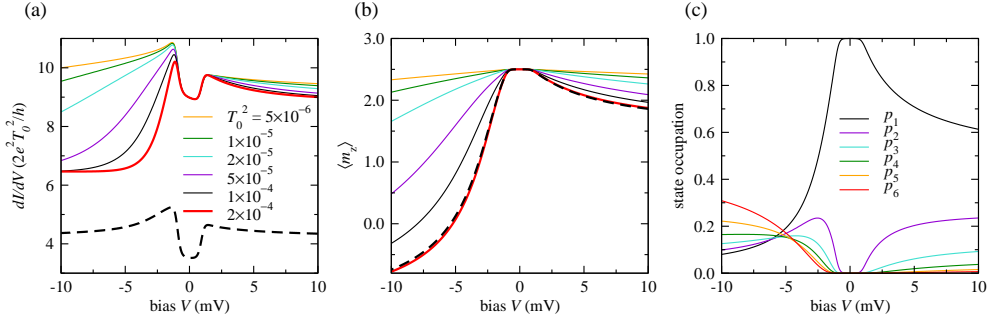}
	\caption{Simulated tunneling spectra of a Mn atom on Cu$_2$N with spin 
$S=5/2$ at different coupling strengths $T_0^2$ between tip and sample and at a 
magnetic field of $B_z=7$~T. Simulation parameters are $g=1.9$, 
$D=-40\ \mu$eV, $E=0$, $U=1.3$, $J\rho_s=-0.02$, $T=1$~K, and a tip 
polarization of $\eta^{\rm t}=0.3$.
\textbf{(a)} Normalized spectra, and \textbf{(b)} averaged magnetic moment 
$\left\langle m_z \right\rangle$ in units of $\hbar$ for $T_0^2$ ranging from 
$5\times 10^{-6}$ to $2\times 10^{-4}$, corresponding to a tunneling resistance 
of $R_T\approx270$~M$\Omega$ to $\approx6.8$~M$\Omega$ or a 
stabilization current of $I\approx 35$~pA -- $1.5$~nA at a bias of 
$eV=+10$~meV. The dashed lines in (a) and (b) are calculated with $U=0$ at 
$T_0^2=2\times 10^{-4}$.
\textbf{(c)} 
Occupancy of the six eigenstates at steady-state conditions for $T_0^2=2\times 
10^{-4}$.}
\label{fig:Pumping_Mn}
\end{figure}

This behavior changes drastically when a spin-polarized tip is used. We have 
seen that, in particular for spin systems with a potential scattering 
$U=S/2$, the tunneling conductance and thereby the scattering rate depends 
strongly on the bias polarity (Figure~\ref{fig:2nd-order}). Experimentally, this 
was first detected for Mn adatoms adsorbed on Cu$_2$N \cite{Loth10} and 
successfully discussed theoretically in a second order perturbative scattering 
model \cite{Delgado10, Delgado10a}. Figure \ref{fig:Pumping_Mn} shows the 
simulated spectra of a Mn spin system at high field which -- due to the small 
magnetic anisotropy -- leads to an almost equidistant energy difference between 
the five eigenstates. 
These eigenstates are well described by pure 
states with the magnetic quantum numbers $m_z=-5/2, -3/2,...,+5/2$.	

At low tip sample couplings the $dI/dV$ spectrum is similar to the ones 
simulated without rate-equations (Figure~\ref{fig:Fe}d and e), but at higher 
coupling we observe a drastic reduction of the differential conductance at 
negative bias. This decrease is concomitant with the reduction of the 
average magnetization $\langle m_z \rangle = \tr \left( \hat{S}_z \sum_i p_i 
\left| \psi_i \right\rangle\left\langle \psi_i\right|\right)$. For the highest 
$T_0$ the average magnetization becomes even negative at negative bias showing 
clearly the inversion of the state populations. Interestingly, the strong bias 
asymmetry in the $dI/dV$ spectrum is only due to the potential scattering. 
Simulating the system with $U=0$ results in a much less asymmetric spectrum, 
even though the bias dependence of the state populations and average 
magnetization are unaltered.

Finally, we have to remark that experimentally an effective interaction of the 
substrate conduction electrons with the Mn spin of $G_{S}=2.7\ \mu$S was found 
\cite{Loth10}. 
This is much higher than the spin-substrate scattering determined solely by the 
spectroscopically estimated $J\rho_0\approx-0.02$ which results to 
$G_{S}=2e^2/h\times(J\rho_s)^2S(S+1)\approx 0.3\ \mu$S. 
This means that 
in this particular system roughly 90\% of the spin relaxations with the 
substrate electrons must originate from scattering processes which do not 
directly leave their fingerprint in the observable differential conductance 
spectrum. 
In this context, ab-inito density functional calculations have found an about  
$3.1$ times higher coupling to the substrate via the $3d_{xz}$ and 
$3d_{x^2-y^2}$ orbitals than experimentally observed \cite{Novaes10}, while the 
approach discussed here neglect any orbital symmetry.

\subsection{Current induced correlations}

\begin{figure}[tbp]
\centering
\includegraphics[width=\textwidth]{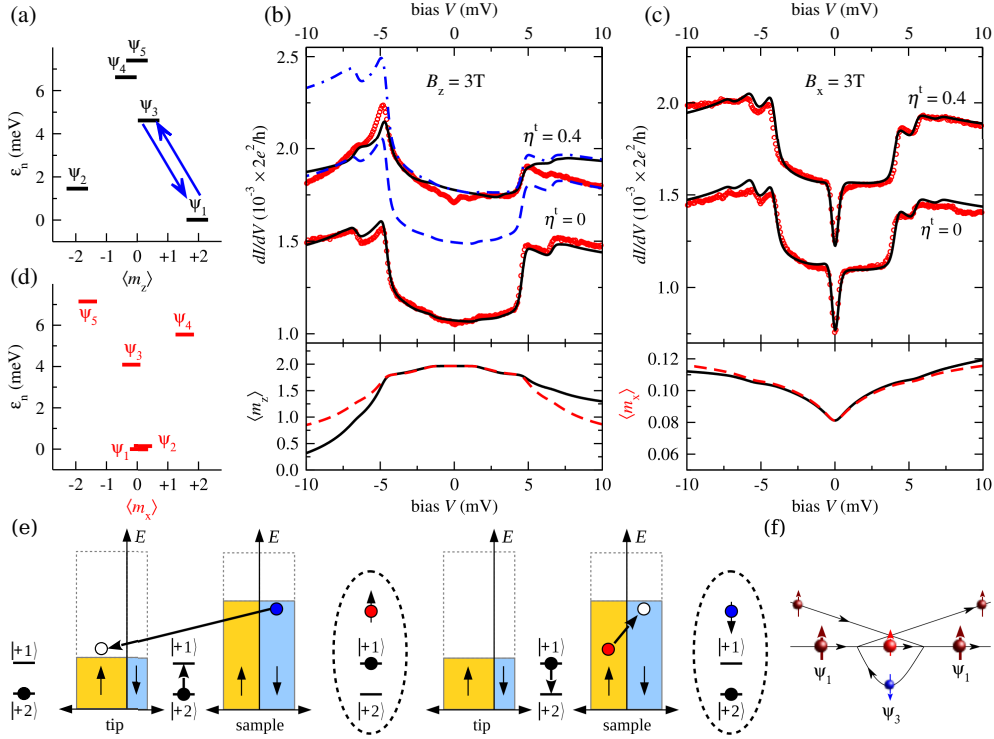}
	\caption{Correlations induced by tunneling substrate electrons.
\textbf{(a)} Schematic state diagram for an Fe atom on Cu$_2$N at 
$B_z=3$~T. The blue arrows illustrate the third order transition (131). 
\textbf{(b)} and \textbf{(c)} Top panels: Experimental data from reference 
\cite{Loth10b} of a Fe atom on Cu$_2$N measured with a spin averaging tip 
($\eta^{\rm t}=0$) and a spin polarized tip ($\eta^{\rm t}\approx0.4$) at 
applied field in $z$ (a) and $x$ (b) direction (colored circles). The best fits 
(full lines) results in $g=2.11$, $D=-1.6$~meV, $E=0.31$~meV, $U=0.35$, 
$J\rho_s=-0.1$, $T_{\rm eff}=1$~K, and $T_0^2=3.2\times10^{-4}$. Spectra 
obtained and simulated with spin-polarized tips are shifted 
vertically by $0.5\times10^{-3}(2e^2/h)$. The dashed (dotted dashed) line in 
(b) is calculated with $U=0$ (in zero-current approximation).
Bottom panels: Corresponding average magnetic moment in field direction in 
units of $\hbar$ (full lines: spin-polarized tip, dashed 
lines: spin-average tip). \textbf{(d)} Schematic state diagram 
for an Fe atom at $B_x=3$~T. \textbf{(e)} and \textbf{(f)}  Schematic 
illustration (e) and Feyman-like diagram (f) of the third order scattering 
process (131) which leads to the peak at $V\approx-5$~mV in panel (a).}
\label{fig:Pumping_SP_Fe}
\end{figure}

Comparing the dynamical model outlined above with experimental data that require
the evaluation up to third order scattering is a very intriguing test of the 
capability, as well as the limitations, of this approach. Thus, we will now 
analyze spectroscopic data that have been measured for Fe adatoms on Cu$_2$N 
using spin-averaged and spin-polarized tips \cite{Loth10b}. Figure 
\ref{fig:Pumping_SP_Fe} shows the experimental data obtained for two different 
magnetic field directions; 
along the main anisotropy axis (easy axis, $B_z$) which leads to a strong 
polarization of the eigenstates along the magnetic field 
(Figure~\ref{fig:Pumping_SP_Fe}a) and, perpendicular to it, along the hard axis 
($B_x$) which produces only a small polarization of the lowest eigenstates 
(Figure~\ref{fig:Pumping_SP_Fe}d).

In this experiment the spin polarization was deliberately changed by vertical 
manipulation (''picking-up``) \cite{Eigler90, Loth10} of a Mn atom onto the apex 
of the tip, changing its polarization from $\eta^{\rm t}\approx0$ to 
$\eta^{\rm t}\approx0.4$ and enabling the study of the identical atom with and 
without a polarized tip. The experimental data for the spin-average 
and for the spin-polarized measurements along the hard axis can be well 
simulated within one set of parameters (figure~\ref{fig:Pumping_SP_Fe}c). 
Surprisingly, the spectra are almost identical to simulations using only the 
zero-current approximation. Thus, the moderate coupling between tip and sample 
does not disturb significantly the state populations. This is also evident by 
plotting the average magnetic moment along the applied field (lower panels in 
figure~\ref{fig:Pumping_SP_Fe}b and c). The small magnetic field (compared to 
the anisotropy energy) of only $B_x=3$~T along the magnetically hard-axis cannot 
polarize the Fe atom significantly (Figure~\ref{fig:Pumping_SP_Fe}d). This 
means, that the spin imbalance produced by a spin-polarized current has only 
little influence on the $dI/dV$ spectrum. However, note that 
the small shift towards higher energies of the step at $|V|\approx4$~mV, which
has its origin in transitions between the states $\left|\psi_1\right\rangle$ 
and $\left|\psi_3\right\rangle$, might be due to an interaction between the 
spin system on the surface and the spin polarized electrode of the tip 
\cite{Baumgaertel11, Misiorny13}.

The situation changes quite drastically when the magnetic field is applied 
along the easy axis and the spectrum is taken with the spin-polarized 
tip (Figure~\ref{fig:Pumping_SP_Fe}b). A strong peak at $V\approx-5$~mV appears 
concomitant with the apparent disappearance of the step at precisely the 
same energy, that was clearly visible in the spectrum measured with the 
spin-averaging tip. Fortunately, the model describes this 
spectrum quite well, too and thus allows us to discuss the physical 
origin of these differences. 

Comparing the spectrum with the one calculated in the 
zero-current-approximation reveals that the disappearance of the step is indeed 
due to the increased coupling between tip and sample. The origin lies in the 
finite $U$, quite similar to the differential conductance reduction observed at 
negative bias for the Mn system (Figure~\ref{fig:Pumping_Mn}). Indeed, we see 
that the spectrum calculated with the same set of parameters except $U=0$ leads 
to a spectrum in which the step-like increase of the $dI/dV$ at $V<-4$~mV is 
maintained. Furthermore, we notice that the calculated magnetization 
$\left\langle m_z \right\rangle$ decreased abruptly below this tunneling bias 
voltage, much faster than for the spin-averaging tip, due to the increased 
population of the states $\left|\psi_2\right\rangle$ 
and $\left|\psi_3\right\rangle$ (Figure~\ref{fig:Pumping_SP_Fe}b lower panel).

While the effects discussed above give a plausible explanation for the overall 
shape of the spectrum, there is still a noticeable discrepancy between 
experiment and simulation. In particular, the experimental $dI/dV$ data show a 
significantly stronger peak at 
$V\approx -5$~mV than one would expect based on the simulation, resembling of 
the effects of strong correlations discussed in section~\ref{sec:limit} (which 
are not covered by the model). Thus, it is worth to dissect the characteristic 
peak which has its 
origin in the third order process (131) in which the spin system is scattered 
from 
$\left|\psi_1\right\rangle\rightarrow\left|\psi_3\right\rangle\rightarrow\left|
\psi_1\right\rangle$. With the spin-polarized tip this transition has a 
significantly higher probability to occur when the electrons tunnel from sample 
to tip. The process is illustrated in figures~\ref{fig:Pumping_SP_Fe}d and 
\ref{fig:Pumping_SP_Fe}f. 
It starts with an $\left|\downarrow\right\rangle$ electron 
in the sample which scatters on the spin system changing its grounstate from 
$\left|\psi_1\right\rangle$ to $\left|\psi_3\right\rangle$ and then tunnels, 
with increased probability, as $\left|\uparrow\right\rangle$ electron into the 
majority states of the tip. The spin-down hole in the sample will lead to a 
slightly higher spin-up electron density close to the Fe atom. Next, a 
$\left|\uparrow\right\rangle$ electron from the sample fills the 
$\left|\downarrow\right\rangle$ hole 
together with changing the localized spin-state from 
$\left|\psi_3\right\rangle$ back to $\left|\psi_1\right\rangle$ 
(Figure~\ref{fig:Pumping_SP_Fe}d). During 
this process the substrate electrons are correlated with the spin 
state. While the magnetic moment for $\left|\psi_1\right\rangle$ is 
approximately $m_z\approx+2$ and for $\left|\psi_3\right\rangle$ 
$m_z\approx+1$, respectively, at low enough temperature \emph{and} at 
sufficiently negative bias and tunneling current, a 
correlated out-of-equilibrium Kondo-state develops which has the total 
wavefunction
\begin{equation}
\Psi^{total}=\frac{1}{\sqrt{2}}\left(
\left|\downarrow^{\rm s}\right\rangle\left|+2\right\rangle-
\left|\uparrow^{\rm s}\right\rangle\left|+1\right\rangle
\right).
\label{equ:Bias-Kondo-state}
\end{equation}
Even though the experimental data only give hints that this correlated state 
has formed, a similar non-equilibrium Kondo formation has been found in 
transport measurements on carbon nanotubes \cite{Paaske06}. 
The physical properties of such a non-equilibrium correlated state are very 
intriguing \cite{Rosch03, Paaske04}. Entering 
this state, the substrate electrons partly screen the magnetic moment of the Fe 
spin reducing it to $S-1/2$. Such an underscreened Kondo state should show 
a very particular temperature and energy dependence \cite{Coleman03, 
Borda09, Misiorny12, Misiorny12a}. Interestingly, in a system like the one 
discussed here, the transition between weak and strong coupling is not only 
governed by a characteristic temperature $T_K$ \cite{Pletyukhov12}, but the 
formation of this correlated state is deliberately tuned by drastically 
increasing the probability of its creation when applying a spin-polarized 
current of sufficient energy by the probing tip. This enables us to envision 
for example pump-probe experiments \cite{Loth10a} in which the time-evolution of 
the formation and the decay of this correlated state is measured in detail.

\section{Summary}
\label{sec:summary}

In this manuscript I have shown that applying perturbation theory to quantum 
spin systems enables one to describe experimentally measured differential 
conductance spectra with very high accuracy. This enables one to 
obtain a profound understanding of the physical processes on play and to 
separate single- as well as many-electron effects. 

The versatility of low-temperature scanning tunneling measurements on single 
and complexly coupled spin systems led me to believe that we should expect 
a multitude of exciting new experiments for the future, which will further 
deepening our fundamental knowledge on quantum systems in general 
and, in particular, quantum magnetism. 
Perturbative models, like the one outlined here, might be of help in such 
systems. For that, the supplemental material to this manuscript include an 
easy usable software package that allows not only to simulate the differential 
conductance spectra of arbitrary complex spin systems, but additionally 
allows one to fit experimental data to the model.

In this manuscript we have restricted ourselves to the adiabatic limit and 
neglected any time dependence in the parameters. However, it is straightforward 
to expand this model to capture also time dependent pump-probe measurements. 
In such a framework, coherent state superpositions could be accounted for by 
using, for example, a Bloch-Redfield approach in which the spin system is 
coupled to an open quantum-system \cite{Argyres64, Cohen89, Timm08} and in 
which interactions up to third order are included to account not only for the 
decay but also for the creation of coherences. 

Additionally, the study of the coupling between the spin-system and other 
(quasi)-particles, like phonons, photons, or magnons would be very 
interesting. Furthermore, while the perturbative approach fails when the system 
enters the strong-coupling Kondo-regime, it would be very intriguing to combine 
the simple model outlined here, with exact quantum impurity models.

%

%
%
%

%
%
%
%
%
%
%
%
%
%
%
%
%
%
%
%
%
%
%
%
%
%
%
%
%
%


\ack

I thank Peter Wahl for invaluable discussions about the original 
Appelbaum paper. Furthermore, I am very thankful for fruitful discussions and 
for providing me with experimental data to Christian Ast, Katharina Franke, 
Andreas Heinrich, Cyrus Hirjibehedin, Alexander Khajetoorians, Nicolas 
Lorente, Sebastian Loth, Alexander Otte, and Marteen Wegewijs. This work was 
supported by the SFB 767.


\appendix

\section{Matrix elements for arbitrary spin polarization}
\label{ap:1}

The two arbitrary spin density matrices $\varrho^{\rm t}$ and $\varrho^{\rm s}$ 
are a full description of the ensemble states in the tip and sample electron 
bath close to the Fermi energy. The eigenvectors 
$\left|\varphi_i^{\rm 
t}\right\rangle$ and $\left|\varphi_i^{\rm 
s}\right\rangle$ of $\varrho^{\rm t}$ and $\varrho^{\rm s}$ 
are representative eigenstates of these \textit{incoherent} ensembles and 
enables one to calculate the interaction transition intensities between them as:
\begin{equation}
\xi^{x,y,z,I}_{i'f'}=
\sqrt{\lambda_{i'}^{\rm t}\lambda_{f'}^{\rm s}}
\left\langle\varphi_{f'}^{\rm s}\right|
\hat{\sigma}_{x,y,z,I}
\left|\varphi_{i'}^{\rm t}\right\rangle,
\label{eq:xi}
\end{equation}
with $\lambda_i^{\rm t}$ and $\lambda_i^{\rm s}$ as the eigenvalues of the 
corresponding eigenvectors. Note, that due to the in general incoherent spin 
ensembles in tip and sample equation \ref{eq:xi} has to be evaluated 
independently for all combinations of $i'=1,2$ and $f'=1,2$.

For the localized spin system we can write similarly the 
transition intensities between the eigenstates:
\begin{equation}
\Xi_{if}^{x,y,z}=\left\langle\psi_{f}\right|
\hat{S}_{x,y,z}\left|\psi_{i}\right\rangle,
\end{equation}
so that the absolute square of the matrix element in equation 
\ref{eq:Matrix1} calculates to:
\begin{equation}
\left|\mathcal{M}_{(1)if}^{\rm 
t \rightarrow s}\right|^2=\sum_{i',f'}\left|\frac12\left(\xi^{x}_{ i'f' } \Xi^ {
x } _ { if }+\xi^{y}_{i'f'}\Xi^{y}_{if}+\xi^{z}_{i'f'}\Xi^{z}_{if}\right)
+U\xi^{I}_{i'f'}\delta_{if}\right|^2.
\label{eq:Matrix-appendix}
\end{equation}

The interference term between $\mathcal{M}_1$ and $\mathcal{M}_2$ 
(equation~\ref{eq:Kondo_U_M}) can be evaluated in an analogous way 
leading to:
\begin{eqnarray}
M_{fi}\tilde{M}_{mf}M_{im}=\sum_{\genfrac{}{}{0pt}{}{j,k,l=}{\{x,y,z,I\}}}\,
\sum_ { i',m',f'}
 \bar{\xi}^{l}_{f'i'}\Xi^{l}_{fi}\times
 \tilde{\xi}^{k}_{m'f'}\Xi^{k}_{mf}\times
\xi^{j}_{ i'm' }\Xi^{j}_{im},
\end{eqnarray}
and for the reversed order processes:
\begin{eqnarray}
 M_{fi}M_{mf}\tilde{M}_{im}=\sum_{\genfrac{}{}{0pt}{}{j,k,l=}{\{x,y,z,I\}}}\,
\sum_ { i',m',f'}
 \bar{\xi}^{l}_{f'i'}\Xi^{l}_{fi}\times
 \xi^{k}_{m'f'}\Xi^{k}_{mf}\times
\tilde{\xi}^{j}_{ i'm' }\Xi^{j}_{im}.
\end{eqnarray}
The tilde ($\tilde{\xi}$) and bar ($\bar{\xi}=\xi^\dagger$) account for 
processes which starts and ends in the sample, and for processes in 
which tunneling electrons traverse the junction in the opposite direction, 
i.\,e.\ which starts in the sample and ends in the tip, respectively. They can 
be calculated in an analogous way with equation \ref{eq:xi}.

\section*{References}
\bibliographystyle{unsrt}

\end{document}